\documentclass[12pt,review,sort&compress]{elsarticle}

\usepackage{amssymb}
 \usepackage{lineno}
\usepackage{xcolor}
\usepackage{hyperref}
 \usepackage{pdflscape}

\newcommand{\highlight}{} 

\makeatletter
\def\ps@pprintTitle{%
  \let\@oddhead\@empty
  \let\@evenhead\@empty
  \def\@oddfoot{\reset@font\hfil\thepage\hfil}
  \let\@evenfoot\@oddfoot
}
\makeatother

\begin{document}

\begin{frontmatter}

\title{The impact of Covid-19 vaccination in Aotearoa New Zealand: a modelling study}

\author[inst1]{Samik Datta}
\author[inst2,inst3]{Giorgia Vattiato}
\author[inst4]{Oliver J. Maclaren}
\author[inst5]{Ning Hua}
\author[inst6,inst7]{Andrew Sporle}
\author[inst2]{Michael J. Plank\corref{cor1}}

\cortext[cor1]{Corresponding author email address: michael.plank@canterbury.ac.nz}

\affiliation[inst1]{organization={Population Modelling group, National Institute of Water and Atmospheric Research},
                   city={Wellington},
                   country={New Zealand}}

\affiliation[inst2]{organization={School of Mathematics and Statistics, University of Canterbury},
                   city={Christchurch},
                   country={New Zealand}}

\affiliation[inst3]{organization={Manaaki Whenua},
                   city={Lincoln},
                   country={New Zealand}}

\affiliation[inst4]{organization={Department of Engineering Science, University of Auckland},
                   city={Auckland},
                   country={New Zealand}}
                   
\affiliation[inst5]{organization={Precision Driven Health},
                   city={Auckland},
                   country={New Zealand}}

\affiliation[inst6]{organization={Department of Statistics, University of Auckland},
                   city={Auckland},
                   country={New Zealand}}
                   
\affiliation[inst7]{organization={iNZight Analytics Ltd.},
                   city={Auckland},
                   country={New Zealand}}
                   
\date{}

\begin{abstract}
Aotearoa New Zealand implemented a Covid-19 elimination strategy in 2020 and 2021, which enabled a large majority of the population to be vaccinated before being exposed to the virus. This strategy delivered one of the lowest pandemic mortality rates in the world. However, quantitative estimates of the population-level health benefits of vaccination are lacking. Here, we use a validated mathematical model of Covid-19 in New Zealand to investigate counterfactual scenarios with differing levels of vaccine coverage in different age and ethnicity groups. The model builds on earlier research by adding age- and time-dependent case ascertainment, the effect of antiviral medications, improved hospitalisation rate estimates, and the impact of relaxing control measures. The model was used for scenario analysis and policy advice for the New Zealand Government in 2022 and 2023. We compare the number of Covid-19 hospitalisations, deaths, and years of life lost in each counterfactual scenario to a baseline scenario that is fitted to epidemiological data between January 2022 and June 2023. Our results estimate that vaccines saved 6650 (95\% credible interval [4424, 10180]) lives, and prevented 74500 [51000, 115400] years of life lost and 45100 [34400, 55600] hospitalisations during this 18-month period. Making the same comparison before the benefit of antiviral medications is accounted for, the estimated number of lives saved by vaccines increases to 7604 [5080, 11942]. Due to inequities in the vaccine rollout, vaccination rates among M\=aori were lower than in people of European ethnicity. Our results show that, if vaccination rates had been equitable, an estimated 11--26\% of the 292 M\=aori Covid-19 deaths that were recorded in this time period could have been prevented. We conclude that Covid-19 vaccination greatly reduced health burden in New Zealand and that equity needs to be a key focus of future vaccination programmes.

\end{abstract}


\newpageafter{author}

\end{frontmatter}

\clearpage

\section{Introduction}
Aotearoa New Zealand used a combination of border and community control measures to minimise transmission of SARS-CoV-2 until high vaccine coverage could be achieved \cite{baker2020successful,vattiato2022assessment}. Prior to the introduction of the B.1.1.529 (Omicron) variant into the community in January 2022, New Zealand had recorded only around 2.5 confirmed community cases per 1000 people and 0.01 Covid-19 deaths per 1000 people \cite{mohgithub}. By this time, 77\% of the population (90\% of those aged over 12 years) had received at least two doses of the Pfizer/BioNTech BNT162b2 vaccine and 27\% of the population (35\% of adults) had received a third dose. By 1 April 2022, third dose coverage had increased to 51\% of the population (66\% of adults). Children aged 5 to 11 years became eligible for vaccination on 17 January 2022 and by 1 April, 54\% of this age group had received at least one dose and 17\% had received two doses \cite{mohgithub}.

Following transmission from cases in managed isolation facilities into the community in January 2022 \cite{douglas2022tracing} and subsequent relaxation of border controls, New Zealand experienced a series of Omicron waves. Between 1 January 2022 and 30 June 2023, there were around 465 confirmed cases, 5.5 hospitalisations, and 0.61 Covid-19 deaths per 1000 people \cite{mohgithub}. Aotearoa New Zealand's cumulative excess all-cause mortality rate, up to the end of 2022, has been estimated as 0.215 per 1000 people \cite{kung2023New}, one of the lowest pandemic excess mortality rates in the world.  

Achieving high primary series and booster dose coverage before widespread transmission occurred was a key pillar of New Zealand's pandemic response.
Despite good vaccination rates overall, coverage was lower for M\=aori, partly because of inequities in the vaccine rollout \cite{whitehead2021will,whitehead2022structural,waitangi2021haumaru}. On 1 February 2022, primary series coverage was 76\% and 88\% for M\=aori adults under and over 65 years old, respectively. This compares to 88\% and 93\% for people of European ethnicity (the largest ethnicity group in Aotearoa New Zealand). In addition, M\=aori are at higher risk of severe illness and death due to Covid-19 \cite{steyn2020estimated,steyn2021maori,whitehead2023inequities}. The age-standardised Covid-19 mortality rate and hospitalisation rate in 2022-23 were approximately 75\% higher for M\=aori than for people of European ethnicity (see Supplementary Material sec. S10).

Globally, it has been estimated that vaccines saved 19--20 million lives in their first year of use \cite{watson2022global}. However, international estimates cannot simply be mapped onto Aotearoa New Zealand's population because of its elimination strategy and unique epidemic trajectory.
Comparing outcomes with other countries that also pursued an elimination strategy, but experienced waves of transmission with lower vaccine coverage (e.g. Hong Kong \cite{smith2022covid,xie2023resurgence}), suggests that vaccination saved many lives. However, quantitative estimates of the reduction in health burden due to vaccination in New Zealand are currently lacking. 

Here, we use a validated mathematical model of SARS-CoV-2 transmission dynamics and health impact in New Zealand to estimate the number of hospitalisations and deaths that were prevented by vaccines between January 2022 and June 2023. {\highlight This corresponds to the time period in which the Omicron family of variants was dominant \cite{esr2023genomics}.} The model uses robust estimates for the effectiveness of the Pfizer vaccine against infection, severe disease and death caused by Omicron variants \cite{khoury2021neutralizing,cromer2023neutralising,andrews2022covid}. The model builds on our earlier work \cite{lustig2023modelling} by adding several new features. These include age- and time-dependent case ascertainment, the effect of antiviral medications on fatality rates, improved estimates of the age-dependent infection hospitalisation ratio, the effect of relaxing control measures on contact rates, and the impact of new Omicron sub-variants. The model was used for health system planning, scenario analysis and policy advice for the New Zealand Government in 2022 and 2023.

We fit the model to data on Covid-19 cases, hospitalisations and deaths between January 2022 and June 2023.
We then consider counterfactual scenarios in which public health measures, behavioural patterns and resulting time-dependent contact rates were the same as the fitted baseline (factual) scenario, but vaccine coverage was lower. 
Our results are important for estimating the population-level health benefits of Covid-19 vaccination, supporting evaluation of the cost-effectiveness of Aotearoa New Zealand's Covid-19 vaccination programme, and for quantifying the impact of inequitable vaccination coverage between population groups. Our findings may be useful in informing preparedness and response to future respiratory pandemic threats.

\section{Methods}

\subsection*{Mathematical model}
\begin{figure}
    \centering
    \includegraphics[trim={2cm 0 2cm 0},clip,width=\linewidth]{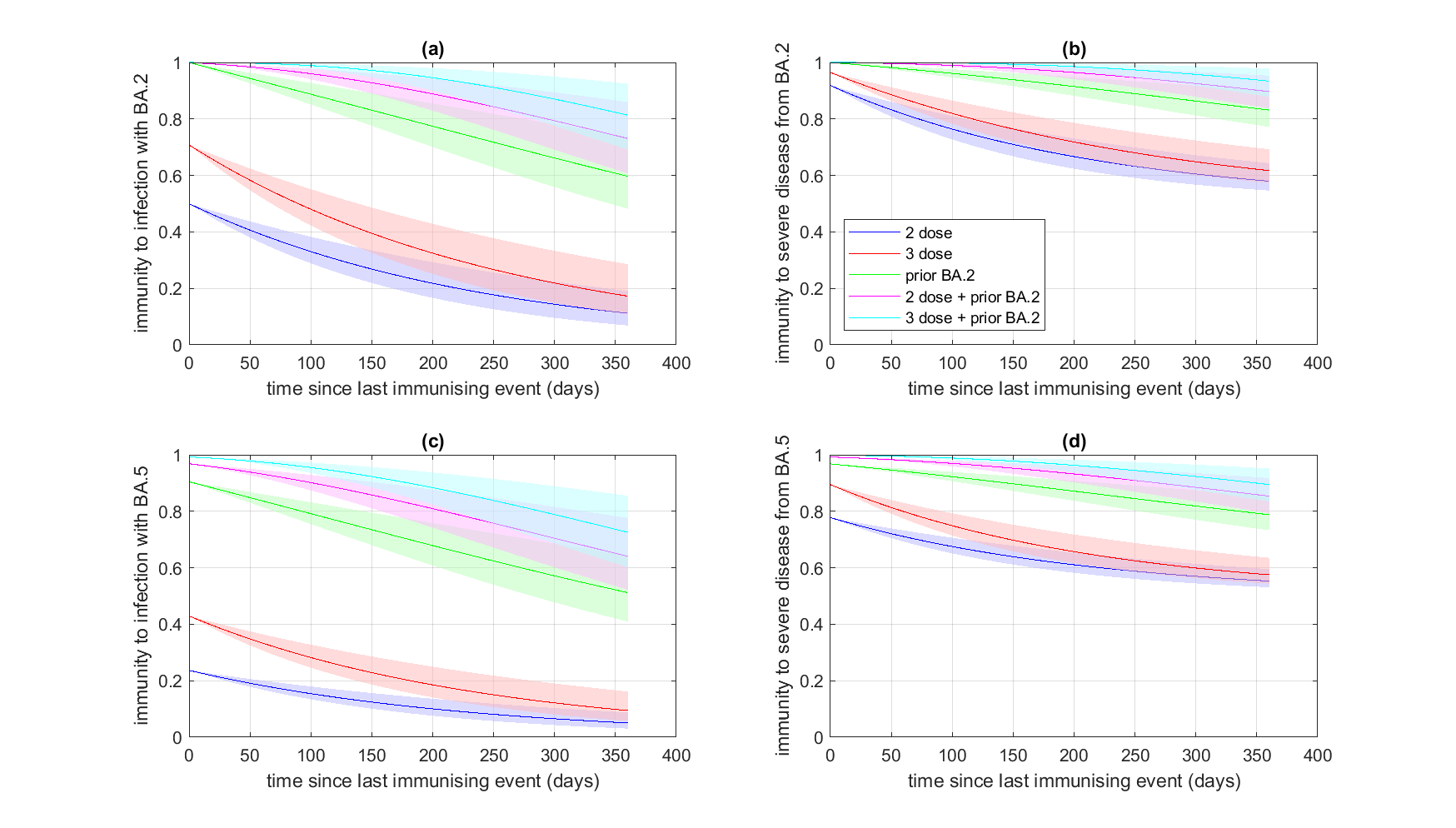}
    \caption{Average immunity under assumed parameter priors against: (a) infection with BA.2, (b) severe disease or death from BA.2, (c) infection with BA.5 and (d) severe disease or death from BA.5 as a function of time since most recent immunizing event. Graphs show immunity following two doses (blue), three doses (red), 0/1 doses and prior infection with BA.2 (green), two doses and prior infection with BA.2 (magenta) and three doses and prior infection with BA.2 (cyan). Immunity from two or more prior Omicron infections also follows the cyan trajectory, regardless of the vaccination status. Immunity against BA.5 derived from prior infection with BA.5 is assumed to follow the same curves as for immunity against BA.2 derived from prior infection with BA.2. Curves are the median and shaded areas are the 5th and 95th percentiles of 500 random draws from the prior.}
    \label{fig:immunity_curves}
\end{figure}

We used an age-structured compartment-based model that includes the effects of vaccination, waning immunity and changes in community contact rates over time (see Supplementary Material sec. S1--S4). {\highlight The population is stratified into 16 five-year age groups.} The model includes different levels of immunity to infection and immunity to severe disease and death. Immunity levels depend on the number of vaccine doses received, prior infection status, dominant SARS-CoV-2 variant, and time since last immunising event (see Figure \ref{fig:immunity_curves}). Parameters governing these immunity curves were taken from \cite{lustig2023modelling}, based on the model of \cite{khoury2021neutralizing,cromer2022neutralising} for the
relationship between immunity and neutralizing antibody titre. 

The primary series and booster rollout in Aotearoa New Zealand almost exclusively used the Pfizer/BioNTech BNT162b2 vaccine and so we do not attempt to model different effectiveness of different vaccine types. {\highlight A vaccine pass system was introduced in December 2021 as part of the Covid-19 Protection Framework \cite{nzgovt2021covid}. Under this system, businesses and venues were required to make proof of vaccination a condition of entry, or to operate with reduced capacity limits. Vaccine mandates were also were introduced for health, disability and education workers and other specified public sector roles \cite{nzgovt2021covidb}. The vaccine pass system was ended in April 2022 and employment-related vaccine mandates were phased out betwen April and September 2022.}   

Vaccination rates in the model were set according to Ministry of Health data on the number of 1st, 2nd, 3rd and 4th or subsequent doses given in each age group between 19 February 2021 and 6 June 2023 (see Supplementary Figure S2). {\highlight At the start of the study period in January 2022, everyone aged over 12 years was eligible for two doses. Eligibility for a third dose was initially limited to over 65-year-olds, but was extended to all adults from 1 February 2022 and  to all over-16-year-olds from 7 April 2022. Children aged 5-11 years became eligible for a two-dose paediatric vaccine series from 17 January 2022. In June 2022, over-50-year-olds became eligible for a fourth dose. From April 2023, all over-30-year olds became eligible for an additional dose 6 months after their more recent dose. In all cases, people outside the age criteria are eligible if they have a specified health condition.} In April 2023, the Pfizer bivalent BA.4/5 vaccine replaced the original formulation for use as a booster dose. However, since this was designed primarily to counter the effects of antigenic evolution, we assume this had the same effectiveness parameters as the original vaccine did against earlier Omicron strains. 

The model has previously been fitted to Aotearoa New Zealand data on Covid-19 cases, hospital admissions and deaths, and used to model the impact of the BA.5 subvariant that caused a wave in July 2022 \cite{lustig2023modelling}. The model used in this study additionally included age- and time-dependent case ascertainment and an adjusted age-dependent infection-hospitalisation ratio to better match the observed age structure in hospital admissions (see Supplementary Material sec. S8). We also made additions to the model to account for changes in the public health response and the impact of new Omicron sub-variants described below. 

On 13 September 2022, the New Zealand Government ended the Covid-19 Protection Framework \cite{nzgovt2021covid}, which meant that mask requirements in public settings such as retail and public transport were lifted, and isolation requirements for household contacts were ended. To model the effect of this policy change, we included an increase in contact rates in the model, with the magnitude of the effect fitted to subsequent data (see Supplementary Material sec. S6). This is in addition to the increase in contact rates and within-age-group mixing in March-April 2022, representing relaxation of non-pharmaceutical interventions and associated precautionary behaviour, as previously modelled \cite{lustig2023modelling}.

Around the same time the Covid-19 Protection Framework ended, the eligibility criteria for antiviral treatments for Covid-19 were broadened to include everyone aged over 65 years, M\=aori and Pacific people aged over 50 years, and people with specified comorbidities. This was followed by a significant increase in the number of prescriptions for antivirals (Supplementary Figure S5). To model this, we assumed that the infection-fatality ratio at time $t$ was a linearly decreasing function of the proportion of notified cases at time $t$ who filled a prescription for either Paxlovid or molnupiravir, the two Covid-19 antiviral treatments in widespread use in New Zealand during this time period (see Supplementary Material sec. S8). The effect size was fitted to subsequent data. We did not attempt to distinguish between different effect sizes for different antivirals as the data were insufficient for this. We did not model any effect of antivirals on the infection-hospitalisation rate, partly because in some cases antiviral prescriptions were only filled on or after hospital admission.   

In November 2022, a mixture of Omicron lineages (most notably CH.1.1 and BQ.1.1) sharing similar sets of genetic mutations displaced BA.5 as the most common variant \cite{esr2023genomics}. To model the growth advantage of these subvariants, we assumed that, on 15 November 2022, a new combination of functionally equivalent immune evasive subvariants became dominant. This is a highly simplified model: we did not attempt to model these subvariants individually, but instead assumed that their net effect can be captured via a reduction in the level of population immunity. We implemented this via the same model mechanism as for the BA.5 subvariant that displaced BA.2 as the dominant variant in July 2022 \cite{lustig2023modelling}, with an effect size parameter that was manually calibrated using subsequent data (see Supplementary Material, sec. S7). 

We fitted the model to data on confirmed Covid-19 cases, hospitalisations and deaths between 1 February 2022 and 13 August 2023. Hospitalisations were defined to be those categorised by the Ministry of Health as receiving hospital treatment for Covid-19. Deaths were defined to be those where the cause of death was classified as ``COVID underlying'' or ``COVID contributory''. The Ministry of Health also publishes the number of deaths that occurred within 28 days of a positive Covid-19 test. The definition that we use is a more accurate estimate of the impact of Covid-19 as it excludes incidental deaths (i.e. those where the cause of death was found to be not Covid-19-related).  Fitting was done with an approximate Bayesian computation (ABC) rejection algorithm {\highlight on a subset of model parameters including time- and age-dependent contact rates, testing and clinical severity parameters, waning rate, and antiviral effect size. This results in a set of 150 model realisations, with each one representing a combination of parameter values sampled from the approximate joint posterior distribution. For all results, we report 95\% credible intervals (CrI) corresponding to the 2.5th and 97.5th percentiles of the sample (see Supplementary Material sec. S9 for details).} 

For the main analysis, we used the Ministry of Health's Health Service User (HSU) dataset (see Supplementary Table S2) for population size by age and prioritised ethnicity. However, this dataset underestimates the size of the M\=aori population and has a different age profile than official population counts and estimates \cite{statsnz_hsu}. We therefore also ran a sensitivity analysis using StatsNZ's annual population estimates \cite{tepou2021dhb}, in which the M\=aori population size is 11\% larger than in the HSU data (see Supplementary Material sec. S5 for details).

\subsection*{Counterfactual scenarios}

\begin{figure}
    \centering
    \includegraphics[width=\textwidth, trim = 2.5cm 0 2.5cm 0, clip]{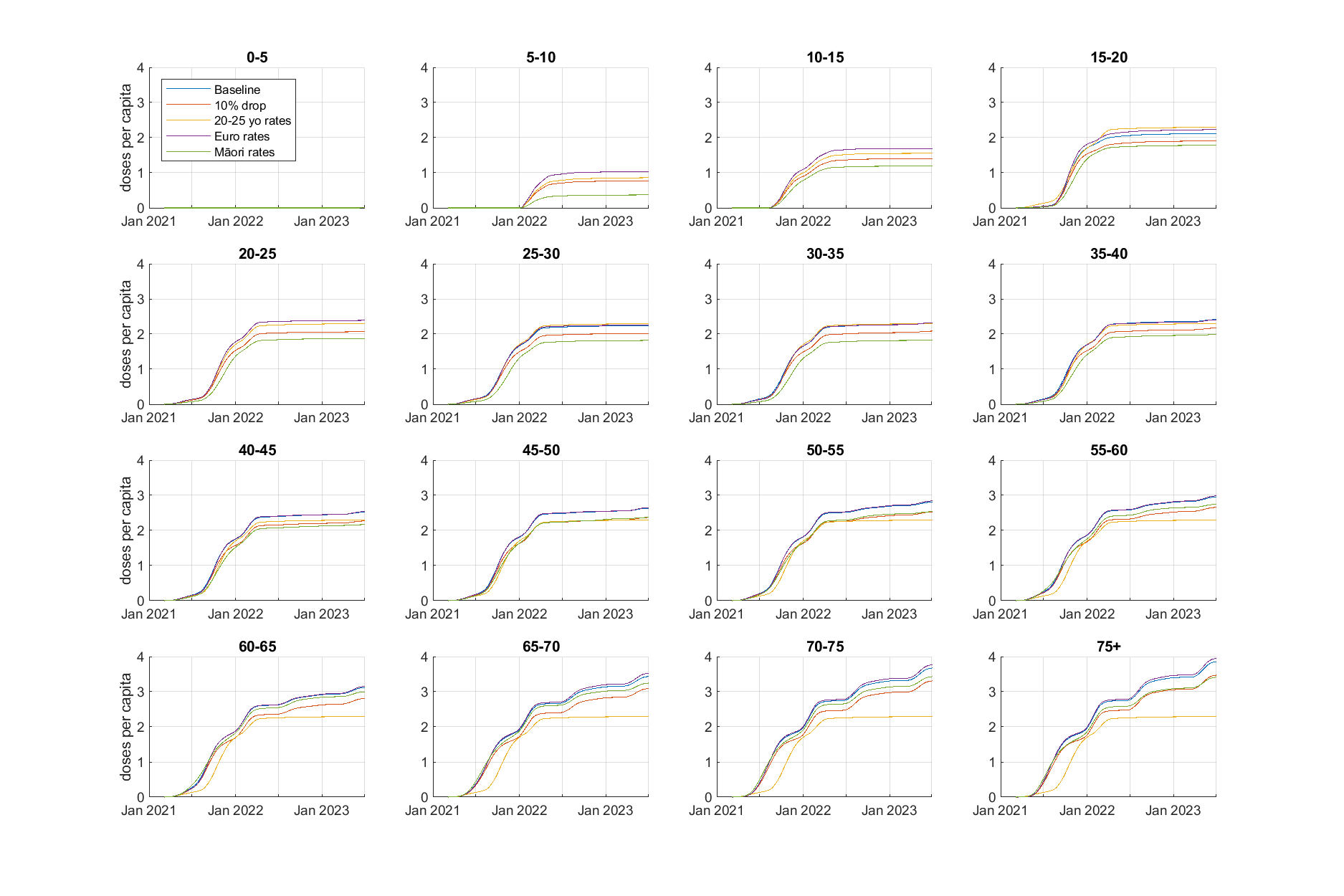}
    \caption{Cumulative number of vaccine doses per capita in the baseline scenario (representing actual data up to 6 June 2023, blue) and counterfactual scenarios: vaccination rates set to a proportion $p_v=0.9$ of actual vaccination rates at all ages (red); vaccination rates at all ages set to actual vaccination rates in the 20-25-year-old group (yellow); vaccination rates set to actual vaccination rates for the European/other (purple) and M\=aori (green) ethnicities. The scenarios with no vaccination and no vaccination in under-60-year-olds are not shown as they are either zero or equal to the baseline (blue) curves. Each panel shows a different age group. Graphs only show total number of doses for ease of visualisation but in the model these are broken down into 1st, 2nd, 3rd, and 4th or subsequent doses. }
    \label{fig:vaccine_counterfactuals}
\end{figure}

We ran model simulations with the same set of parameter combinations produced by the baseline fitting procedure, but with the number of daily $d^\mathrm{th}$ vaccine doses $v_{id}(t)$ in age group $i$ at time $t$ modified to represent a given counterfactual scenario. For each accepted parameter combination, we considered the following counterfactual scenarios: (1) no vaccination; (2) vaccination rates set to a proportion $p_v$ of actual vaccination rates at all ages; (3) no vaccination of under-60-year-olds; (4) vaccination rates for all adults set to actual vaccination rates in the 20-25-year-old group (resulting in a reduction in coverage particularly for older adults); (5-6) vaccination rates set to actual vaccination rates for European/other and M\=aori ethnicities respectively at all ages (see Figure \ref{fig:vaccine_counterfactuals}). Where vaccination rates were changed by a multiplicative factor, or set equal to those of another age or ethnicity group, this change was applied to the time series of daily 1st, 2nd, 3rd, and 4th or subsequent doses (see Supplementary Table S3 for details).  We also ran the baseline scenario and scenario (1) but with no antiviral medications in the model. This was to enable a secondary comparison of outcomes with and without vaccination before the impact of antivirals on the death rate is accounted for. 

In each scenario, we assumed that the time-dependent contact rates, mixing between age groups, timing and characteristics of new variants were the same as in the fitted factual scenario. In each scenario, we calculated total number of infections, hospital admissions, deaths and years of life lost (YLL) due to Covid-19, and the peak hospital occupancy.

Population-level YLL were calculated using New Zealand cohort life tables published by StatsNZ \cite{statsnz_cohort_life_tables}. The life tables provide life expectancy estimates by age, sex and year of birth. We used the model output for the total number of Covid-19 deaths in each five-year age group and imputed the number of model deaths in one-year age groups and by sex according to the distribution of actual Covid-19 deaths. We then multiplied this by the life expectancy estimate for people of that age and sex in 2022, and calculated total YLL by summing across all age groups. We only calculated YLL at the population level and did not attempt to calculate YLL for specific ethnicity groups. 

{\highlight To estimate the number of M\=aori hospitalisations and deaths that could have been prevented if M\=aori vaccination rates had been equal to those of European/other ethnicity, we first calculated the model death (or hospitalisation) rates in age group $i$ in the scenarios with European/other vaccination rates ($E_i$) and M\=aori vaccination rates ($M_i$) aggregated over the modelled time period. We then applied the relative difference between these rates to the actual number of M\=aori deaths (or hospitalisations) that occurred in each age group in the same time period, and summed over all age groups:
\[
    \textrm{preventable outcomes} = \sum_i \left(1-\frac{E_i}{M_i}\right) \left( \textrm{actual M\=aori outcomes in age group } i \right)
\]
}

The time period considered for counterfactual scenarios was 1 January 2022 to 30 June 2023, {\highlight which is the first 18 months of the Omicron period.}  We did not consider hospitalisations and deaths prior to 1 January 2022 and did not attempt to model the dynamics of the Delta outbreak that occurred in Auckland between August 2022 and January 2023 \cite{plank2022using}. The national vaccine rollout was still underway during the period and intensive non-pharmaceutical interventions were used to suppress transmission and contain the outbreak to the Auckland metropolitan area up to December 2022. Over 99\% of all cases and around 98\% of all deaths reported up to 30 June 2023 occurred after 1 January 2022.

Data and fully documented code to reproduce the results in this article is publicly available at \url{https://github.com/SamikDatta/covid19_vaccination/tree/main}.

\section{Results}

\subsection*{Baseline model fit}

\begin{figure}
    \centering
    \includegraphics[width=\linewidth, trim = 2cm 0 2cm 0, clip]{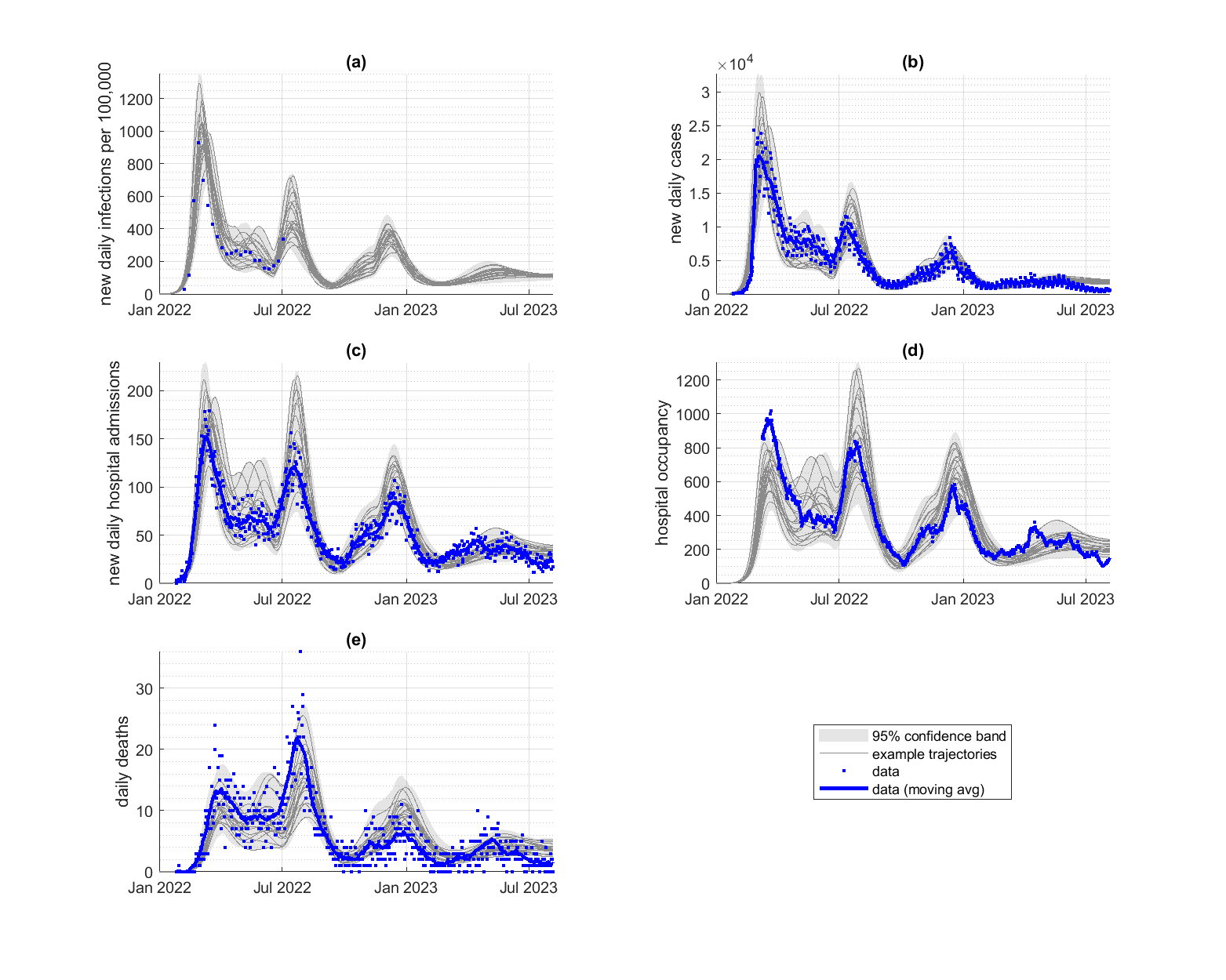}
    \caption{Results for the baseline (factual) scenario showing: (a) new daily infections per 10,000 people; (b) new daily cases; (c) new daily hospital admissions; (d) hospital occupancy; (e) daily deaths. Graphs show the curvewise 95\% CrI (grey shaded area) and a random sample of model trajectories (solid gray lines). Model was fitted to data (blue points) up to 13 August 2023; solid blue curves show the moving average of the data in a 7-day window for cases and hospital occupancy, a 14-day window for admissions, and a 21-day window for deaths. }
    \label{fig:modelresults1}
\end{figure}

The model provided a reasonable fit to the time series of cases, hospitalisations and deaths between January 2022 and June 2023 at the aggregate level (Figure \ref{fig:modelresults1}) and within ten-year age bands (Supplementary Figures S8--S9). This time period included three distinct waves dominated by different Omicron subvariants: the BA.1/BA.2 wave that peaked in March 2022, the BA.5 wave that peaked in July 2022, and the predominantly CH.1.1/BQ.1.1 wave that peaked in December 2022. There was also a smaller wave that peaked in April 2023 and was dominated by a mixture of recombinant XBB lineages \cite{esr2023genomics}. However, unlike the two preceding waves, this wave did not require specific model assumptions about a more transmissible or immune evasive variant, but was largely captured by the default model assumptions about continuous background waning of immunity.  

The fitted baseline model had 3163 (95\% CrI [2169, 4561]) deaths and 28800 [22200, 35600] hospital admissions between 1 January 2022 and 30 June 2023. The actual numbers for this period were 3196 deaths and 28763 admissions. Our model estimated that Covid-19 was responsible for 39900 [27600, 57300] YLL in this time period, an average of approximately 13 YLL per Covid-19 fatality.

\subsection*{Counterfactual scenarios}

In the counterfactual scenario with no vaccination, there were a total of 9953 [6616, 14824] deaths, 115100 [79500, 174800] YLL, and 74300 [56500, 90900] hospital admissions between 1 January 2022 and 30 June 2023 (see Table \ref{tab:results}). This represents an additional 6650 [4424, 10180] deaths, 74500 [51000, 115400] YLL, and 45100 [34400, 55600] admissions relative to the baseline scenario (see Supplementary Table S5). An alternative basis for estimating the effect of vaccination is to compare model outcomes with and without vaccination in a scenario with no antivirals. Under this comparison, the scenario with no vaccination had an additional 7604 [5080, 11942] deaths and 82400 [56100, 129300] YLL (see Supplementary Table S5).  

\begin{figure}
    \centering
    \includegraphics[trim={0 0 0 1.05cm},clip,width=\linewidth]{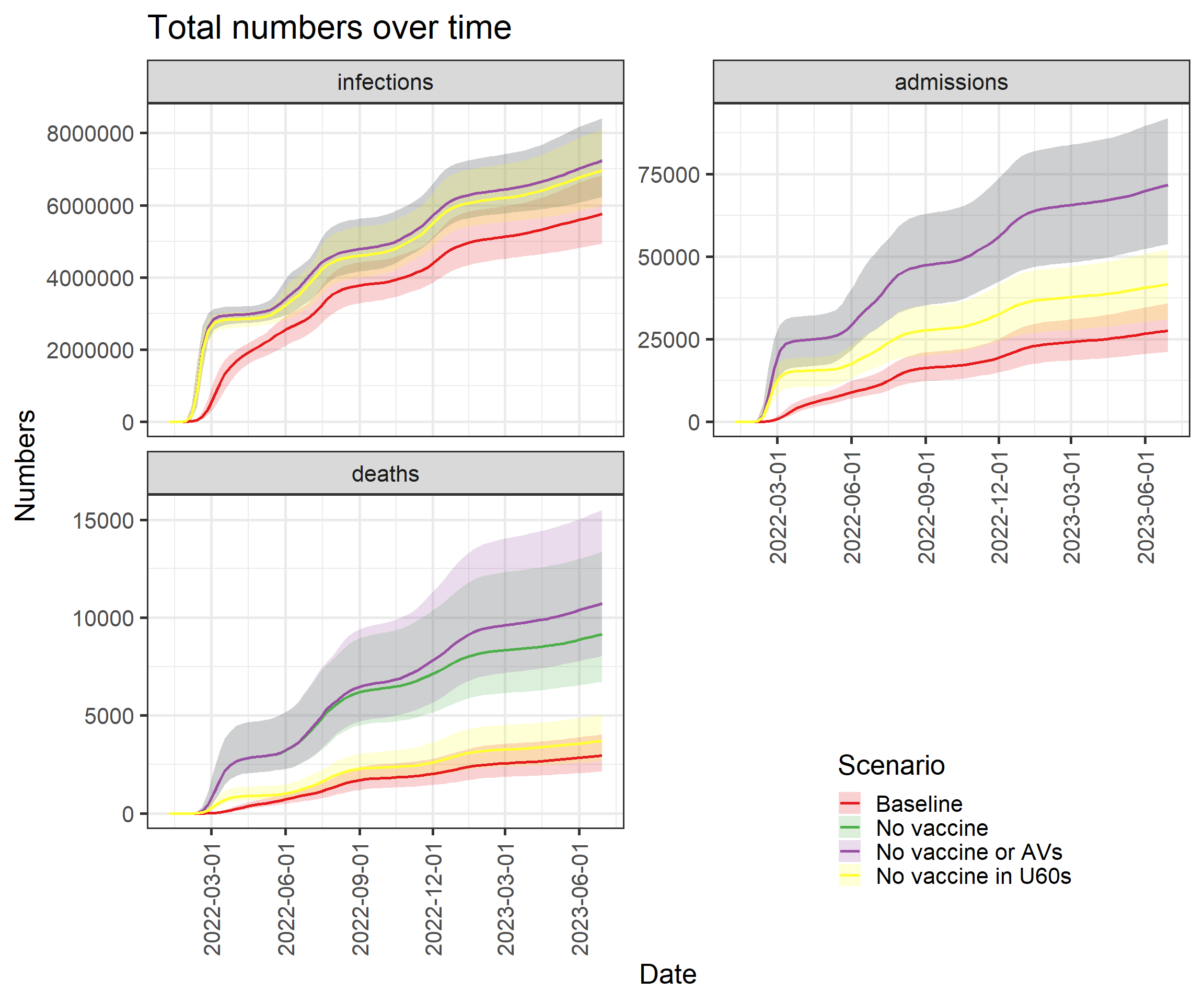}
    \caption{Cumulative number of infections, hospital admissions and deaths over time in the baseline scenario (red) and the scenarios with no vaccination (green), no vaccination or antivirals (purple), and no vaccination of under 60-year-olds (yellow). Graphs show the median (solid curves) and 95\% CrI  (shaded areas) for each scenario. Results for the other scenarios considered are not shown here as they are relatively close the baseline scenario (see Supplementary Figure S10). Note that, for infections and admissions, the results for the `No vaccine' and `No vaccine or antivirals' scenarios are identical as the model assumes that antivirals only affect the fatality rate.}
        \label{fig:comparison}
\end{figure}

These additional health impacts were concentrated in the first Omicron wave in February-March 2022 (see Figure \ref{fig:comparison}), which infected around 50\% more people than in the baseline scenario and had higher average severity since all infections were in immune naive individuals. Peak hospital occupancy in this wave with no vaccines was 5542 [3695, 7526]. This is around 7 times higher than in the baseline scenario, a level that would have completely overwhelmed hospital capacity (see Discussion). Additional health impacts continued to accumulate over the remainder of the study period, albeit at a slower rate (Figure \ref{fig:comparison}).   

In the no vaccination scenario, the attack rate (proportion of the population with at least one infection) in the over-75-year-old group was 52\% [43\%, 62\%]. This was significantly lower than attack rates in younger groups, which were over 90\% for under-60-year-olds. This means that, in the no vaccination scenario, there was still as a significant immune naive population in elderly groups at the end of the simulation period. As a consequence, the health burden prevented by vaccination will continue to increase over time.  

Scenarios (2)--(4) had varying levels of impact intermediate between the baseline and no vaccination scenarios (Table \ref{tab:results}). Of these, scenario (3) with no vaccination in under-60-year-olds had the biggest impacts, most notably on hospital admissions and YLL. This is due to shallower age gradient in hospitalisation risk relative to fatality risk (see Supplementary Table S2) and an increase in fatalities under-60-year-olds where life expectancy is higher.

\begin{landscape}
\begin{table}
\small
    \centering
    \begin{tabular}{p{4.1cm}p{2.8cm}p{2.9cm}p{3.2cm}p{3cm}p{3cm}}
    \hline
     Scenario    & Infections \hspace{5mm} (millions) & Admissions \hspace{5mm} (thousands)  & Deaths  & YLL \hspace{12mm} (thousands) & Peak occupancy \\
     \hline
{\em Actual}  & -  & $28.8$ & $3196$   & $33.9^*$ & $1016^{**}$   \\
     \hline
{\em Model scenarios} \\
(0) Baseline & 5.83 [5.12, 6.77] & 28.8 [22.2, 35.6] & 3163 [2169, 4561] & 39.9 [27.6, 57.3] & 833 [549, 1266]  \\ 
(0a) No AVs & 5.83 [5.12, 6.77] & 28.8 [22.2, 35.6] & 3818 [2576, 5427] & 45.3 [31.0, 64.8] & 833 [549, 1266]  \\ 
(1) No vaccine & 7.27 [6.52, 8.29] & 74.3 [56.5, 90.9] & 9953 [6616, 14824] & 115.1 [79.5, 174.8] & 5542 [3965, 7256]  \\ 
(1a) No vaccine or AVs & 7.27 [6.52, 8.29] & 74.3 [56.5, 90.9] & 11457 [7671, 17341] & 128.4 [88.4, 194.0] & 5542 [3965, 7256]  \\ 
(2) 10\% drop in rates & 5.96 [5.25, 6.91] & 33.1 [25.6, 40.7] & 3788 [2569, 5519] & 46.8 [32.2, 68.2] & 1036 [693, 1388]  \\ 
(3) No vaccine in U60s & 7.02 [6.27, 8.01] & 43.8 [33.2, 52.5] & 3937 [2706, 5695] & 59.5 [41.0, 87.6] & 2845 [2076, 3620]  \\ 
(4) 20-25-year-old rates & 6.03 [5.29, 7.00] & 33.2 [25.6, 41.0] & 3906 [2660, 5666] & 47.2 [32.5, 68.0] & 916 [593, 1313]  \\ 
(5) Euro/other rates & 5.80 [5.09, 6.74] & 28.4 [21.8, 35.1] & 3100 [2127, 4460] & 39.2 [27.2, 56.4] & 817 [531, 1260]  \\ 
(6) M\=aori rates & 6.06 [5.33, 7.02] & 31.9 [24.7, 39.4] & 3522 [2410, 5107] & 43.7 [30.3, 63.4] & 1011 [681, 1312]  \\ 

    \hline
    \end{tabular}
    \caption{\small Model results (median and 95\% CrI) in each scenario for the total number of infections, hospital admissions, deaths, and years of life lost (YLL), and the peak hospital occupancy, between 1 January 2022 and 30 June 2023.  Scenarios are: (0) baseline (actual vaccination rates); (0a) no antivirals (actual vaccination rates); (1) no vaccination; (1a) no vaccination or antivirals; (2) vaccination rates set to a proportion $p_v=0.9$ of actual vaccination rates at all ages; (3) no vaccination of under-60-year-olds; (4) vaccination rates at all ages set to actual vaccination rates in the 20-25-year-old group; (5-6) vaccination rates set to actual vaccination rates for European/other and M\=aori ethnicities respectively at all ages. $^*$Estimated from actual deaths using cohort life tables via the same method as for model YLL calculations. $^{**}$Includes some incidental hospitalisations (i.e. patients who were positive for Covid-19 but not receiving treatment for Covid-19).  }
    \label{tab:results}
\end{table}
\end{landscape}

Scenario (2), which was a 10\% reduction in vaccination rates at all ages, and scenario (4), in which vaccination rates at all ages were set to those of 20-25-year-olds had very similar outcomes. The reduction in vaccination in older groups in scenario (4) was partially offset by the fact that 20-25-year-olds generally received their primary series and third dose somewhat later than older groups (see Figure \ref{fig:vaccine_counterfactuals}). The timing of the first wave in February-March 2022 therefore meant that older groups were less affected by waning immunity in scenario (4) than in the baseline scenario. With different timing, the additional health impact in this scenario could have been even larger, emphasising the importance of achieving high vaccine coverage in older age groups. 

Relative differences among scenarios in the number of infections were smaller ($<$25\% above baseline in all scenarios). This shows that, over the 18-month time period considered, the primary benefit of vaccination in the model was to reduce the risk of severe disease and death rather than to prevent transmission. The relative differences in hospital admissions and deaths of all scenarios compared to baseline are shown in Figure \ref{fig:comparison}.

Scenario (5) in which European/other vaccination rates were applied to the whole population had 64 [43, 91] fewer deaths and 500 [400, 600] fewer hospitalisations than baseline (Figure \ref{fig:comparison} and Supplementary Table S5). Scenario (6) in which M\=aori vaccination rates were applied to the whole population had a death rate that was 8.0 [5.4 12.2] per 100,000 higher and a hospitalisation rate 68.7 [52.1, 82.7] per 100,000 higher than scenario (5). These differences account for 13--29\% and 14--23\% of the observed differences in age-standardised Covid-19 mortality and hospitalisation rates respectively (see Supplementary Table S4). 

Applying the relative differences in age-specific death and hospitalisation rates between scenarios (5) and (6) to the actual number of M\=aori deaths and hospitalisations suggests that if M\=aori vaccine uptake had been at the same level as European/other, 32 [30, 36] deaths and 457 [412, 505] hospitalisations could have been prevented.

In the sensitivity analysis using StatsNZ projections instead of the HSU dataset for population denominators, the results for scenarios (0)--(4)  (Supplementary Table S6) were similar to the main analysis shown in Table \ref{tab:results}. However, scenario (6) in which M\=aori vaccination rates were applied had significantly worse outcomes than in the main analysis. The estimated numbers of preventable M\=aori deaths and hospitalisations in the sensitivity analysis were 77 [72, 84] and 998 [914, 1104] respectively. 

Considering both the main and the sensitivity analysis, our results suggest that 11--26\% of the 292 M\=aori Covid-19 deaths and 10--22\% of the 4616 M\=aori hospitalisations that were recorded during the time period could have been prevented had vaccination rates been equitable.

\section{Discussion}
We have used a validated mechanistic model of Covid-19 transmission dynamics and health impact in Aotearoa New Zealand to explore outcomes in a range of counterfactual scenarios where public health measures and time-varying contact rates were the same as the baseline (factual) scenario, but vaccine coverage was lower. Our results showed that vaccines saved an estimated 4600--9500 lives and prevented 34000--58000 hospitalisations between January 2022 and June 2023, the first 18 months of widespread community transmission. These estimates will continue to increase over time as the model estimated that a significant proportion of the elderly population remained uninfected in June 2023. 

Benefits cannot always be attributed exclusively to a single intervention such as vaccines, particularly when interventions act multiplicatively to reduce risk. In our model, this applies to the benefits of antiviral medications and vaccines. To address this, we compared scenarios with and without vaccination and antivirals. When the comparison of outcomes with and without vaccination was made before the benefit of antivirals on the death rate is taken into account, the estimated number of lives saved by vaccines increased to 5300--10800.  

These estimates only include direct health impacts of Covid-19 and do not account for indirect effects, such as a potential increase in the death rate from Covid-19 or other causes due to an overloaded healthcare system. Similarly, the modelled hospital occupancy does not take account of any change in admission thresholds that may have occurred due to capacity constraints. It should therefore be interpreted as an estimate of the demand for hospital treatment rather than a prediction of realised occupancy.

Vaccination may have prevented other adverse outcomes, such as symptomatic disease leading to attendance at primary care, lost productivity, and incidence of Long Covid \cite{notarte2022impact,tran2023efficacy}. However, we only focused on hospitalisations and deaths and did not consider other outcomes in the model due to a lack of relevant data.  

Adverse events following immunisation with COVID-19 vaccines in Aotearoa New Zealand are monitored by the statutory regulator Medsafe. Detailed reports are available at \cite{medsafe_adverse}. Serious adverse events are extremely rare; however we have not attempted a formal risk-benefit analysis here.   

A global modelling study estimated that vaccines prevented approximately 20 million deaths up to December 2021 \cite{watson2022global}, which equates to around $2.5$ deaths prevented per 1000 primary series doses. Another analysis estimated that vaccines prevented 105,900 deaths in England up to August 2021 \cite{ukhsa2021vaccine}, equivalent to $1.4$ per 1000 primary series doses. By comparison our results equate to between $0.54$ and $1.28$ deaths prevented per 1000 primary series doses (of which there were $8.45$ million during the simulation period). 

There are several reasons why the number of deaths prevented per dose would be lower in the context of our study. In most countries, the initial vaccine rollout prioritised high-risk groups and took place during a period of high global community transmission of the Alpha, Beta and Delta variants. It is to be expected that there are diminishing returns from expanding the rollout to lower-risk groups, even though these are almost certainly still net beneficial. The Omicron variant is less pathogenic than earlier variants of concern \cite{nyberg2022comparative} and effective treatments have improved over time. Finally, although vaccine effectiveness against severe disease caused by Omicron has remained high, effectiveness against transmission is significantly reduced. Globally, as well as in Aotearoa New Zealand, cumulative attack rates were far higher by June 2023 than in December 2021.

Counterfactual modelling is a valuable tool to understand and learn from the potential consequences of alternative scenarios with the benefit of hindsight provided by subsequent data \cite{mishra2021comparing,binny2021early,imai2023quantifying}. However, it has important limitations. 

In reality, in a scenario with low vaccine coverage, the occurrence of higher hospitalisation and death rates would have prompted different policy actions and behavioural responses \cite{gimma2022changes}, which would have altered the transmission dynamics compared to the counterfactual scenarios presented here. Therefore our results should not be interpreted as predictions of what would have happened if Aotearoa New Zealand had not achieved as high a vaccine coverage as it did. Nevertheless, in the absence of vaccination, non-pharmaceutical interventions would only have delayed rather than prevented infections. It is likely that the large majority of unvaccinated individuals would eventually have been infected, meaning a comparable number of hospitalisations and deaths as in the modelled counterfactual would have accrued, albeit over a longer time period. Our results are therefore a valid estimate of the reduction in health burden that is directly attributable to vaccines. We also note that the need for a more stringent policy or behavioural response would have incurred additional direct and indirect costs.

The model for transmission dynamics does not include ethnicity as a variable. In estimating the number of M\=aori deaths that could have been prevented if vaccination rates had been equitable, we assumed that the per capita death rate for a subgroup with a given vaccination rate would be equal to the per capita death rate if that vaccination rate was applied to the whole population. This is a simplistic assumption that ignores the differential effect of vaccination on transmission between and within ethnicities.   
Our estimates apply only to the period from 1 January 2022 to 30 June 2023. They do not include vaccine-preventable health impacts that disproportionately affected M\=aori during the 2021 Delta outbreak \cite{tewhatuora_dashboard,smith2022covid,summers2023improvements}. 

We found that estimates for the number of preventable M\=aori hospitalisations and deaths were larger when using StatsNZ population estimates than when using the HSU data for population denominators. It is recognised that the HSU data underestimates total M\=aori population size and has biases in the age and sex structure as it only includes those with recent engagement with health services \cite{statsnz_hsu}. Therefore the results that use the StatsNZ population estimates may be more accurate for M\=aori impacts. The significant gap between these point to the need for consistent collection of high-quality ethnicity data, accurate population size estimates, and explicit consideration of the uncertainty due to the differences in key data sources.

Our results show that inequitable vaccine coverage is one contributing factor to the higher Covid-19 hospitalisation and mortality rates experienced by M\=aori. Other factors likely include increased exposure to infection due to high rates of employment in high-contact occupations, larger average household size and lower-quality housing \cite{mcleod2020covid,harvey2023summary}, high rates of comorbidities \cite{steyn2020estimated,yon2014cohort,gurney2020inequity}, poor access to healthcare \cite{whitehead2022structural}, and a health system that has historically underserved M\=aori \cite{waitangi2021haumaru}. All these factors are worthy of attention and further study. Including the effects of ethnicity in dynamic epidemiological models is challenging due to the paucity of data and need for modelling assumptions about contact rates between and within ethnic groups. This a key area needing further research in Aotearoa New Zealand.

Our results cannot be used to estimate outcomes if there had been widespread transmission of SARS-CoV-2 in Aotearoa New Zealand earlier in the pandemic when fewer effective treatments were available. Some pre-Omicron variants were more pathogenic \cite{nyberg2022comparative,davies2021increased} and it is likely that health impacts would have been larger if uncontrolled transmission of these variants had occurred in an unvaccinated population. {\highlight We did not attempt to model counterfactual scenarios based on alternative vaccination policies, such as absence of the vaccine pass system and employment-related vaccine mandates. This is because there is no quantitative or qualitative data to inform modelling assumptions about the effect these policies would have had on vaccine update or on contact rates of unvaccinated individuals.}

Our analysis is retrospective and cannot be used to predict the health benefits of future booster rollouts or changes to vaccine eligibility. These will depend on the future prevalence and clinical severity of SARS-CoV-2 variants, the marginal increase in protection offered by additional doses relative to existing levels of hybrid immunity, and how well matched the vaccines are to dominant circulating strains \cite{fabiani2023relative,kirsebom2023effectiveness}.

The uncertainty estimates we have provided arise from the range of parameter values for which model outputs are approximately consistent with epidemiological data in the baseline scenario. There may be additional uncertainty arising from potential model misspecification, omission of relevant mechanisms, or population heterogeneity, which were not accounted for.

Aotearoa New Zealand's Covid-19 pandemic response used a combination of border controls, physical distancing measures including periods of stay-at-home orders, financial support for those unable to work, testing and contact tracing, mass masking, vaccination, and antiviral treatments \cite{baker2020successful,cumming2022going,binny2022real}. This response delivered one of the lowest pandemic mortality rates of any country in the world \cite{cmi2023mortality}. Given that non-pharmaceutical interventions could only delay rather than prevent widespread transmission, it is clear that vaccination was an essential component of this response that saved thousands of lives. Immunity provided by existing vaccine coverage and future vaccination will continue to be an important factor in reducing the Covid-19 health burden. This means the number of lives saved by Covid-19 vaccines in Aotearoa New Zealand will further increase over time. Ensuring equitable access to vaccination and vaccine coverage should be a priority for future public health or pandemic responses.

\subsection*{Acknowledgements}
The authors acknowledge the role of the New Zealand Ministry of Health, StatsNZ, and the Institute of Environmental Science and Research in supplying data in support of this work. The authors are grateful to Fiona Callaghan, Nigel French, Emily Harvey, Markus Luczak-Roesch, Dion O'Neale, Matt Parry, Patricia Priest, the Covid-19 Modelling Government Steering Group and two anonymous reviewers for feedback on earlier versions of this manuscript. The authors acknowledge the contributions of Rachelle Binny, Shaun Hendy, Kannan Ridings, Nicholas Steyn and Leighton Watson to previous models from which this model was developed. This work was partly funded by the New Zealand Department of the Prime Minister and Cabinet and Ministry of Health. 

 \bibliographystyle{elsarticle-num} 

\end{document}


\maketitle


\renewcommand\thesection{S\arabic{section}}
\renewcommand\theequation{S\arabic{equation}}
\renewcommand\thefigure{S\arabic{figure}}
\renewcommand\thetable{S\arabic{table}}

\tableofcontents 

\clearpage

\section{Transmission dynamics} \label{sec:transmission_dynamics}
The susceptible population is divided into $n_A=16$ five-year age groups and $n_S=14$ susceptibility classes depending on the number of vaccine doses received, prior infection status, and the degree of waning (see Figure \ref{fig:diagram}). Infected individuals progress through a series of disease states specific to the susceptibility class from which they were infected. Transitions between compartments are governed by a set of ordinary differential equations for the susceptible ($S$), exposed ($E$), clinical infectious ($I$), subclinical infectious ($A$) and recovered ($R$) compartments for each age group $i=1,\ldots,n_A$ and susceptibility class $k=1,\ldots,n_S$:
\begin{eqnarray}
\frac{dS_{ik}}{dt} &=& -\lambda_i (1-e_{I,k}) S_{ik} + W_{ik} + G_{ik}  \label{eq:dSdt}  \\
\frac{dE_{ik}}{dt} &=& \lambda_i (1-e_{I,k}) S_{ik}  - 1/t_E E_{ik}  \\
\frac{dI_{ik}}{dt} &=& 1/t_E p_{\mathrm{clin},i} (1-e_{S,k}) E_{ik}  - 1/t_I I_{ik}  \\
\frac{dA_{ik}}{dt} &=& 1/t_E \left(1 - p_{\mathrm{clin},i} (1-e_{S,k})\right) E_{ik}  - 1/t_I A_{ik}  \\
\frac{dR_{ik}}{dt} &=& 1/t_I (I_{ik}+A_{ik}) - r_w \hat{r} R_{ik},
\end{eqnarray}
where $e_{O,k}$ is the immunity against outcome $O$ for people in susceptible compartment $k$ (see Sec. \ref{sec:immunity_model}), $t_E$ and $t_I$ are the latent and infectious periods, respectively, $p_{\mathrm{clin},i}$ is the probability of infection causing clinical symptoms in age group $i$, $r_w$ is the waning rate, and $\hat{r}$ is the relative rate of moving from recovered ($R$) to susceptible ($S$). 

For each susceptible compartment, there are associated compartments for people who were infected whilst in that susceptible compartment and are currently: exposed but not yet infectious ($E$); infectious and with clinical symptoms ($I$); infectious and subclinical ($A$); recovered and temporarily immune ($R$). Note that subclinical refers to people who never develop symptoms. For simplicity we do not distinguish between the pre-symptomatic and symptomatic stages of the infectious period for clinical individuals, although it would be possible to do this, for example to model symptom-based interventions. Parameter values are shown in Tables \ref{tab:all_pars}--\ref{tab:age_dep_pars}. The assumed values for the mean latent and infectious periods correspond to a mean generation interval of 3.3 days, which is similar to estimates of \cite{abbott2022estimation,wu2022incubation} for the Omicron variant and shorter than that of previous variants of SARS-CoV-2.

The $W_{ik}$ and $G_{ik}$ terms represent waning and vaccination dynamics (see Sec. \ref{sec:vaccination_and_waning}). The force of infection   $\lambda_i$  acting on age group $i$ is:
\begin{equation}
\lambda_i = \frac{U R_{EI}(t) u_i}{t_I N_i} \sum_{j=1}^{n_A}  M_{ji} \left[\sum_{k=1}^{n_S} (1-e_{T,k}) (I_{jk} + \tau A_{jk}) + t_I n_{\mathrm{seed},j}(t)    \right]
\end{equation}
where $R_{EI}(t)$ is the time-varying reproduction number excluding effects of immunity (see Sec. \ref{sec:time_varying}), $N_i$ is the total population size in each age group, $n_{\mathrm{seed},j}(t)$ is the number of daily seed infections in age group $j$ at time $t$, $\tau$ is the relative infectiousness of subclinical individuals, $u_i$ is the susceptibility of age group $i$ relative to the 60-65 year age group, and $M_{ji}$ is the average number of daily contacts in age group $i$ by someone in age group $j$. The normalising constant $U$ is set to be
\begin{equation*}
U = \rho\left[ \left(p_{\mathrm{clin},j} + \tau(1-p_{\mathrm{clin},j} )\right) u_i M_{ji}  \right]^{-1}    
\end{equation*}
where $\rho[.]$ denotes dominant eigenvalue. This normalisation ensures that the reproduction number at time $t$ would be $R_{EI}(t)$ for a fully susceptible population. $R_{EI}(t)$ represents the value the reproduction number would take if there was no immunity in the population, and hence is unaffected by vaccination, infection and waning dynamics. It therefore provides a way to model time-dependence in contact rates, for example as a result of behavioural change or policy response (see Sec. \ref{sec:time_varying}).

\section{Vaccination and waning}\label{sec:vaccination_and_waning}

\begin{figure}[p!]
    \centering
    \includegraphics[width=0.7\textwidth]{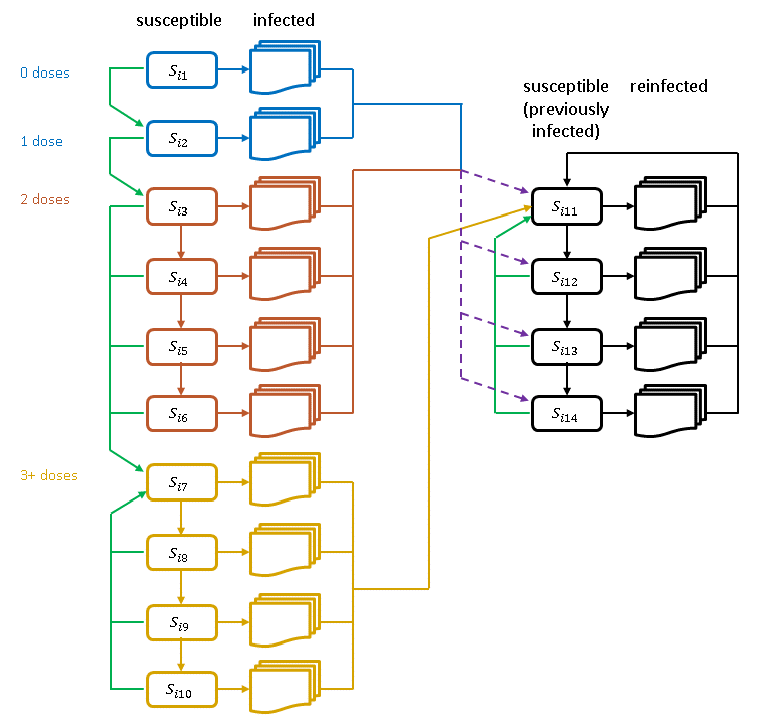}
    \caption{Schematic diagram of the model structure showing the 14 susceptible compartments for age group $i$, indexed as compartments $S_{ik}$ for $k=1,\ldots,14$. Vertical downward arrows represent transition to a susceptible compartment with lower immunity as a result of waning immunity. Green arrows represent transition to a susceptible compartment with higher immunity as a result of vaccination. Horizontal arrows represent infection, which initiates transition through a series of disease states ending in recovery. Following recovery from first infection, individuals who have had at least three vaccine doses (yellow) transition to the highest immunity post-infection compartment $S_{i,11}$; individuals who have had fewer than three vaccine doses (blue and red) transition to a mixture of compartments $S_{i,11}$ to $S_{i,14}$ (dashed purple arrows), representing lower post-infection immunity for these groups. Following recovery from a second or subsequent infection (black), all individuals transition to $S_{i,11}$ regardless of vaccination status.    }
    \label{fig:diagram}
\end{figure}

As indicated above, the $G_{ik}$ term in Eq. \eqref{eq:dSdt} represents transitions between susceptible compartments that occur as a result of vaccination (green arrows in Figure \ref{fig:diagram}). For the purposes of calculating this, we define five groups of susceptible compartments $S^g$:
\begin{eqnarray}
\textrm{0 doses and not previously infected: } \qquad &&  S^g_{i0} = S_{i1}  \\
\textrm{1 dose and not previously infected: } \qquad && S^g_{i1} = S_{i2}  \\
\textrm{2 doses and not previously infected: } \qquad && S^g_{i2} = \sum_{k=3}^6 S_{ik}  \\
\ge\textrm{3 doses and not previously infected: }\qquad  && S^g_{i3} = \sum_{k=7}^{10} S_{ik}  \\
\textrm{                previously infected: } \qquad && S^g_{ip} = \sum_{k=11}^{14} S_{ik}  
\end{eqnarray}
We assumed that all vaccine doses are given to people who are in a susceptible compartment (which is reasonable given the recommendation to wait at least 3 months after testing positive before getting vaccinated). 

\begin{figure}
    \centering
    \includegraphics[width=\linewidth]{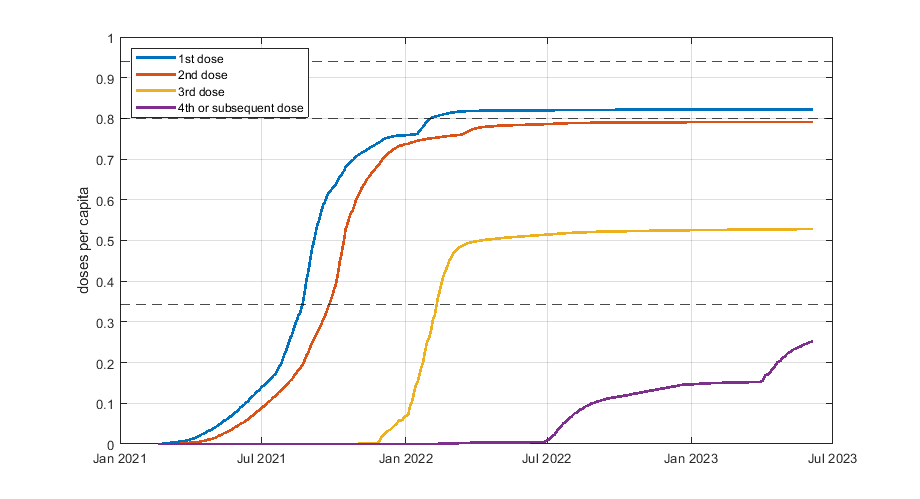}
    \caption{Cumulative number of 1st, 2nd, 3rd and 4th or subsequent vaccine doses as a proportion of Aotearoa New Zealand's total population size, based on Ministry of Health data on actual doses administered up to 6 June 2023 aggregated across all age groups. To assist interpretation, the horizontal dashed lines show (in order from top to bottom), the approximate proportion of the population that is aged over 5 years (eligible for 2 doses as of January 2022), aged over 16 years (eligible for 3 doses as of April 2022), and aged over 50 years (eligible for a 4th dose as of June 2022). Note these are indicative only as they do not include people who were outside the age criteria but who were eligible on the basis of specified health conditions or healthcare worker status. The purple curve combines 4th and subsequent doses and so cannot be interpreted as the proportion of people who have had at least 4 doses. Up to 31 March 2023, this curve is predominantly 4th doses as eligibility for a 5th dose was restricted to small groups; from 1 April 2023 onwards, the curve represents a combination of 4th and 5th doses as the criteria changed at this time so that everyone aged over 30 years became eligible for an additional dose 6 months after their most recent dose. For a age breakdown of vaccination rates see Figure 2 of the main article.  }
    \label{fig:vaccines}
\end{figure}

The total number of people $V_{id}(t)$ in each age group who have received at least $d$ doses of the vaccine at time $t$ is:
\begin{equation} \label{eq:dVdt}
\frac{dV_{id}}{dt} = v_{id}(t)
\end{equation}
where $v_{id}(t)$ is the number of $d^\mathrm{th}$ doses per day given to people in age group $i$ at time $t$ (see Figure \ref{fig:vaccines}).

We assumed that the $v_{id}$ $d^\mathrm{th}$ doses ($d=1,2,3$) given to people in age group $i$ at time $t$ are split pro rata between people who have not been previously infected and people who have. This implies that the daily proportion of those not previously infected in age group $i$ receiving their $d^\mathrm{th}$ dose at time $t$ is 
\begin{equation}
p^u_{i,d} = \frac{v_{i,d}}{V_{i,d-1}-V_{i,d}}
\end{equation}
noting that $V_{i,0}=N_i$, i.e. the total population size in age group $i$. This accounts for $p^u_{i,d}S^g_{i,d-1}$ of the $v_{i,d}$ doses. The remainder of these doses, $v_{i,d}-p^u_{i,d}S^g_{i,d-1}$, are given to previously infected people. This implies that the daily proportion of those previously infected in age group $i$ receiving their $d^\mathrm{th}$ dose at time $t$ is
\begin{equation}
p^p_{i,d} = v_{i,d}\frac{V_{i,d-1}-V_{i,d}-S^g_{i,d-1}}{(V_{i,d-1}-V_{i,d})S^g_{i,p} }
\end{equation}
The corresponding equations for 4th or subsequent doses are
\begin{eqnarray}
p^u_{i,4+} &=& \frac{v_{i,4+}}{V_{i,3}}  \\
p^p_{i,4+} &=& v_{i,4+}\frac{V_{i,3}-S^g_{i,3}}{V_{i,3} S^g_{i,p} }
\end{eqnarray}
We may then write the proportion of compartment $S_{ik}$ receiving a vaccine dose per day as:
\begin{equation}
    P_{i,k} = \left\{ \begin{array}{ll} 
          p^u_{i,1}, & \textrm{if } k=1  \\      
          p^u_{i,2}, & \textrm{if } k=2  \\
          p^u_{i,3}, & \textrm{if } 3\le k \le 6  \\
          p^u_{i,4+}, & \textrm{if } 7\le k \le 10  \\
          \sum_{d=1}^{4+} p^p_{i,d}, & \textrm{if } 11\le k \le 14
          \end{array} \right.
\end{equation}
We assumed that receiving a vaccine dose following prior infection has the effect of moving people back to the first post-infection compartment ($S_{i,11}$) and that receiving a 4th dose without any prior infection has the effect of moving people back to the first 3-dose compartment ($S_{i,7}$) -- see Figure \ref{fig:diagram}.  This is a model simplification to avoid having to keep track of too many different immunity histories

The term $G_{ik}$ appearing in Eq. \eqref{eq:dSdt} is now defined as:
\begin{equation}
    G_{ik} = \sum_{l=1}^{n_S} P_{il} S_{il} Q^V_{lk}
\end{equation}
where $Q^V_{lk}$ is the flux into susceptible compartment $k$ from susceptible compartment $l$ as a result of vaccine doses given to people in susceptible compartment $l$, such that the row sums of the matrix $Q^V$ are all $0$.

The term $W_{ik}$ in Eq. \eqref{eq:dSdt} represents transitions between susceptible compartments and transitions from recovered to susceptible compartments that occur as a result of waning (red, yellow and black arrows in Figure \ref{fig:diagram}) and is defined as: 
\begin{equation} \label{eq:waneTerm}
W_{ik} = r_w \left( \sum_{l=1}^{n_S} S_{il} Q^S_{lk}   +  \hat{r}  \sum_{l=1}^{n_S} R_{il} Q^R_{lk} \right)
\end{equation}
where $Q^S_{lk}$ is the flux into susceptible compartment $k$ from susceptible compartment $l$ (with $Q^S_{kk}\le 0$ representing the flux out of compartment $k$) such that the row sums of the matrix $Q^S$ are all $0$; and $Q^R_{kl}\ge 0$ is the flux into susceptible compartment $k$ from recovered compartment $l$ such that the row sums of $Q^R$ are all $1$.

\section{Immunity model} \label{sec:immunity_model}
The model includes parameters representing the level of immunity against infection ($e_{I,k}$), symptomatic disease ($e_{S,k}$), transmission ($e_{T,k}$), hospitalisation ($e_{H,k}$) and death ($e_{F,k}$) for people in susceptible compartment $k$. In principle, this means there are a total of up to 70 age-independent immunity parameters in the model (14 susceptible compartments times 5 endpoints). To provide a parsimonious parameterisation, we used the model of \cite{khoury2021neutralizing} and \cite{cromer2022neutralising} for the relationship between level of immunity and neutralising antibody titre. The antibody titre is assumed to be a correlate of protection and a given titre is generally more protective against more severe clinical endpoints, in line with the findings of \cite{cromer2023neutralising}. 

For simplicity, we set $e_{Tk}=0$ and $e_{Sk}=e_{Ik}$, i.e. we assumed that immunity reduced the risk of infection but, conditional on infection, did not change the likelihood of symptomatic disease or transmission. We also assumed that immunity against hospitalisation and death never wanes below $e_\mathrm{sev,min}=0.5$. This assumption captured a more durable component of the immune response, for example cellular immunity as opposed to neutralising antibodies, which maintains immunity against severe disease at some baseline long-term level. 
We also assumed immunity for people who are transiently in the one-dose compartment was negligible. Hence $e_{O1}=e_{O2}=0$ for all outcomes $O$ \cite{lustig2023modelling}. 

To model waning immunity, we assumed that the log antibody titre decreases by a fixed amount for each successive susceptible compartment in the same category (i.e. through compartments $k=3,\ldots, 6$, $k=7,\ldots,10$ and $k=11,\ldots,14$). We then mapped the log antibody titre $n_k$ for compartment $k$ to immunity $e_{Ok}$ against outcome $O$ via a logistic function with an outcome-specific midpoint parameter $n_{O,50}$ \cite{khoury2021neutralizing}:
\begin{equation}
    e_{Ok} = \frac{1}{1 + e^{-\kappa(n_k -n_{O,50})} }.
\end{equation}
This framework meant the immunity model could be parameterised with one parameter $n_{l,0}$ for each source of immunity $l$, one parameter for each outcome $O$ and two additional independent parameters: the logistic slope $\kappa$; and the transition rate $r_w$ between successive susceptible compartments, which represents the speed of waning (see Table \ref{tab:all_pars}).

For the post-infection susceptible states, we did not include separate sets of susceptible compartments for people with different vaccination status. Instead, we modelled vaccination-dependent levels of post-infection immunity by moving people to different susceptible compartments dependent on their vaccination status. 
Following recovery from a first infection, we assumed that people who had 3 doses of the vaccine (i.e. those in recovered compartments $k=7,\ldots,10$) all moved initially to the highest immunity compartment $k=11$. This was encoded by the matrix $Q^R$ in Sec. \ref{sec:vaccination_and_waning}: $Q^R_{k,11}=1$ for $k=7,\ldots 10$.

Following recovery from a first infection, we assumed that people who had 2 doses of the vaccine (i.e. those in recovered $k=3,\ldots,6$) moved to one of the lower-immunity compartments $k=12,13,14$ in fixed proportions. To determine what these proportions should be note that, absent any subsequent immunising events, the proportion $q_k(t)$ of a cohort of individuals that entered susceptible compartment $k=11$ at time $t=0$ that is in compartment $k$ at time $t$ satisfies 
\begin{equation}
    \dot{q}_k = \left\{ \begin{array}{ll} -r_w q_k, & k=11, \\ r_w(q_{k-1}-q_k), & k=12,13, \\ r_wq_{k-1}, & k=14, 
    \end{array} \right. 
\end{equation}
where $q_{11}(0)=1$ and $q_k(0)=0$ for $k=12,13,14$. The average log antibody titre of the cohort at time $t$ is $\bar{n}(t)=\sum_k n_k q_k(t)$. 
We therefore set $Q^R_{kl}=q_l(t^*)$ where $t^*$ is such that $\bar{n}(t^*)-\bar{n}(0)=n_{p2d,0}-n_{p3d,0}$, the estimated difference in initial log titre between prior infection plus 2 doses and prior infection plus 3 doses \cite{lustig2023modelling}. 

A similar approach was applied to those moving out of recovered compartments $k=1,2$ (i.e. people with 0 or 1 vaccine doses following recovery from a first infection): we set $Q^R_{kl}=q_l(t^*)$ where $t^*$ is such that $\bar{n}(t^*)-\bar{n}(0)=n_{p,0}-n_{p3d,0}$. 

Following recovery from a second or subsequent infection, we assumed that everyone moved initially to the highest-immunity compartment $k=11$ regardless of vaccination status: $Q^R_{k,11}=1$ for $k=11,\ldots,14$ (see Figure \ref{fig:diagram}).

Immunity from infections that occurred prior to the model seeding date in January 2022 was ignored. This is reasonable as there had only been around 2.5 confirmed community cases per 1000 people up to the end of 2021 \cite{vattiato2022assessment}. Although no representative seroprevalence data is available for New Zealand to estimate the true number of infections, the public health strategy of intensive testing, case finding, contact tracing and source investigation meant that case ascertainment up to the end of 2021 was likely relatively high. Even if case ascertainment during this period were as low as 25\%, that would mean that only around 1\% of the population had been infected prior to the arrival of Omicron and the impact of this on epidemic dynamics during the Omicron period would still be negligible. To ensure the model correctly captured waning of vaccine-derived immunity before the start of the first Omicron wave, we ran the model from a start date of 5 March 2021 (the beginning of the vaccine roll out in New Zealand) with vaccinations administered as per Ministry of Health data, but with no infections prior to the seeding date in January 2022.

\section{Population dynamics} \label{sec:pop_dynamics}
The dynamics of birth, death and ageing are incorporated into the model via additional terms in Eqs. \eqref{eq:dSdt}--\eqref{eq:dVdt} of the form:
\begin{eqnarray}
\frac{dX_{1,k}}{dt} &=& b - r_a X_{1,k} - \mu_1 X_{1,k} \\
\frac{dX_{i,k}}{dt} &=& r_a (X_{i-1,k}-X_{i,k}) - \mu_i X_{i,k} \\
\frac{dX_{n_A,k}}{dt} &=& r_a X_{n_A-1,k} - \mu_{n_A} X_{n_A,k} 
\end{eqnarray}
where $b$ is the birth rate per unit time, $r_a$ is ageing rate per unit time (equal to the reciprocal of the size of the age bands, in this case 5 years) and $\mu_i$ is the per capita death rate per unit time in age group $i$. Here $X$ may be any one of the infection states ($S$, $E$, $I$, $A$, $R$) or $V$. For simplicity we assume that the the aggregate population death rate is independent of the transmission dynamics.

The total number of annual births and the annual death rate in 5-year age bands up to age 75 were taken from StatsNZ data for 2019 \cite{statsnz2022infoshare}. The annual death rate for the over-75-years age group was set to give a similar equilibrium age distribution to the StatsNZ 2022 estimated resident population \cite{statsnz2022infoshare}.

\section{Population size and ethnicity data} \label{sec:pop_data}

For the main analysis, we used the Health Service User (HSU) population dataset supplied by the Ministry of Health in August 2022 (see Table \ref{tab:age_dep_pars}). This contains population size in five-year age groups and for the following prioritised ethnicity groups: M\=aori, Pacific Peoples, Asian, European or Other. The main model only requires total population size in each age group. The age-specific M\=aori and European/other population sizes were used to calculate vaccination rates in the relevant counterfactual scenarios (5 and 6). 

The HSU data includes people who use health services (including births and deaths) in the reference period, or are enrolled in a primary health organisation during the reference period \cite{statsnz_hsu}. In relation to calculating vaccination rates, the HSU dataset has the advantage of consistent reporting of M\=aori and other ethnic groups in the numerator compared with the population denominator. However, it may underestimate the M\=aori population size \cite{statsnz_hsu}. 

We therefore ran a sensitivity analysis using population projections (for 2022) by prioritised ethnicity group, produced by StatsNZ according to assumptions agreed to by the Ministry of Health \cite{tepou2021dhb}. The total population size is very similar in the HSU dataset (5,233,646) and the StatsNZ projection (5,153,500). The total population in each five-year age group is also generally within $\pm 5\%$. However, the M\=aori population size is 11\% larger in the StatsNZ projection (892,130) than in the HSU (801,903), with the discrepancy concentrated in older age groups (see Supplementary Table S7).

\section{Time-varying transmission effects} \label{sec:time_varying}

\begin{figure}[p!]
    \centering
    \includegraphics[width=0.9\linewidth]{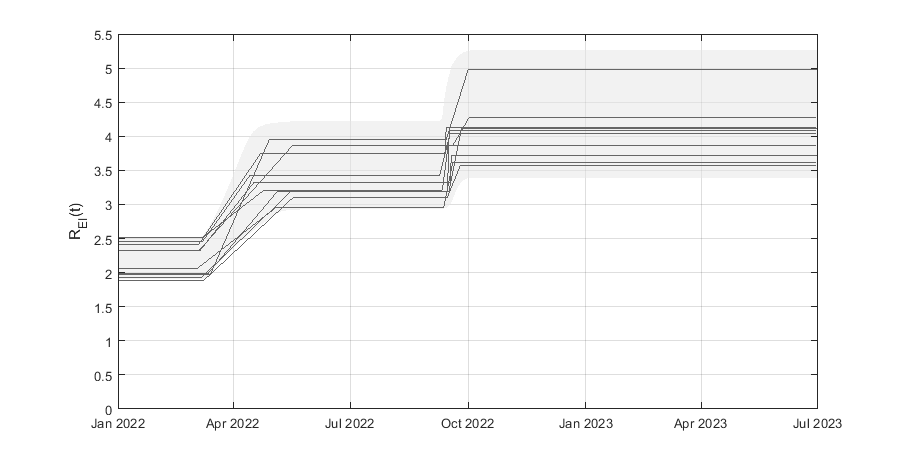}
    \caption{Prior distribution for the time-varying reproduction number excluding immunity $R_{EI}(t)$: ten example time series (solid curves) and 95\% CrI (gray shaded area) calculated from 10,000 random draws of the relevant parameters from the assumed prior (see Table \ref{tab:all_pars}). The graph shows the two transmission ramp-up periods (corresponding to behavioural and/or policy changes) starting in March 2022 and September 2022.}
    \label{fig:control_function}
\end{figure}

 The time-varying function $R_{EI}(t)$ was assumed to increase in two separate phases during 2022, with each change happening as a result of either a behavioural change, a policy change, or a combination of both (see Figure \ref{fig:control_function}). $R_{EI}(t)$ was assumed to increase linearly from $R_{EI,1}\in U[1.9, 2.5]$ to $R_{EI,2}\in U[2.9, 4.3]$ over a 35--75 day time window from the end of period 1 (around 10 March 2022, see Table \ref{tab:all_pars}). This reflects the relaxation of public health measures and associated behaviour change following the peak of the first Omicron wave in early March 2022 (isolation period reduced from 14 days to 10 days on 4 March and to 7 days on 18 March; gathering restrictions eased from 26 March; vaccine pass requirements and most employment-related vaccine mandates lifted from 5 April; gathering restrictions lifted and mask requirements eased from 14 April). $R_{EI}(t)$ was assumed to subsequently increase linearly again from $R_{EI,2}$ by a factor of 10-30\% to $R_{EI,3}$ over a 1--19 day time window from the end of period 2 (around 15 September 2022). This represents a further relaxation of public health measures, with isolation requirements for household contacts and remaining mask requirements lifted from 13 September. The values of $R_{EI,1}$, $R_{EI,2}$, $R_{EI,3}$, as well as the starting dates and time windows over which the transmission increase took place were all fitted with the approximate Bayesian computation (ABC) algorithm as described in Sec. \ref{sec:fitting}.

\begin{figure}[p!]
    \centering
    \includegraphics[width=\textwidth, trim = 2cm 0 2cm 0, clip]{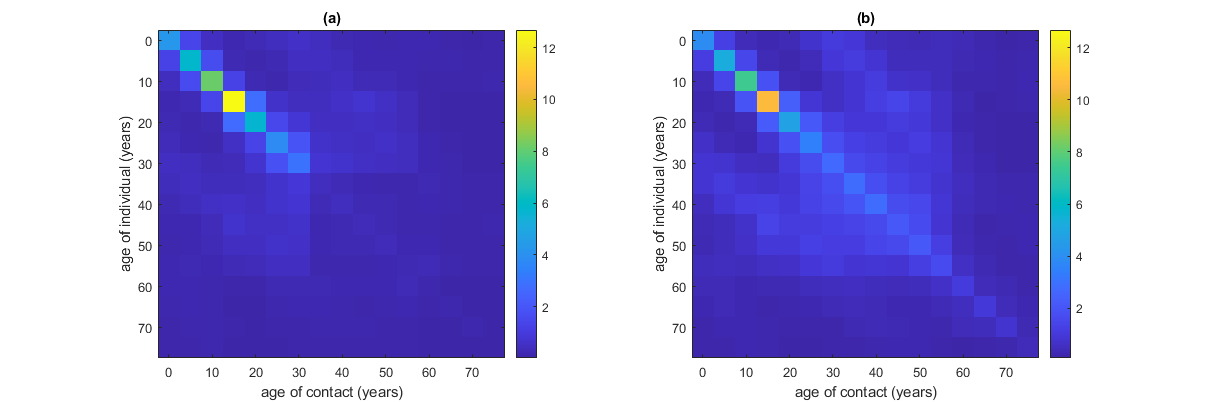}
    \caption[Contact matrices used in the model]{Contact matrices showing the average number of contacts between age groups: (a) during period 1 of the simulation ($M_0$); (b) during periods 2 and 3 of the simulation ($M_1$). The matrix $M_0$ was taken from \cite{vattiato2022modelling} and the matrix $M_1$ was taken from \cite{steyn2022covid} based on \cite{prem2017projecting}. }
    \label{fig:contact_matrices}
\end{figure}

In addition to variations in $R_{EI}(t)$, the contact matrix $M$, defining relative contact rates between age groups, was allowed to vary over time. The contact matrix at time $t$ was defined as
\begin{equation} \label{eq:contact_matrix}
M(t) = (1-\beta(t))M_0 + \beta(t)M_1    
\end{equation}
where $M_1$ is the pre-pandemic contact matrix estimated by \cite{prem2017projecting} and adjusted for the New Zealand population by \cite{steyn2022covid}, and $M_0$ is a modified form of this contact matrix with lower contact rates in older groups and higher contact rates in younger groups (see Figure \ref{fig:contact_matrices}). The matrix $M_0$ was used by \cite{vattiato2022modelling} to model an age-dependent behavioural response and to reproduce the age distribution of cases, admissions and deaths observed in the first Omicron wave in March 2022. The time-varying function $\beta(t)$ was defined to be $0$ in period 1 (i.e. up to around mid-March 2022 - see Table \ref{tab:all_pars}), to increase linearly to $\alpha_M$ over a 50--90 day time window after the end of period 1, and to remain constant at a constant value of $\alpha_M$ subsequently. Thus, Eq. (\ref{eq:contact_matrix}) represents an assumption that contact rates for older groups were significantly reduced in the first Omicron wave in March 2022, but progressively relaxed towards more normal levels between April and June 2022. The value of $\alpha_M\in[0,1]$ was fitted with the ABC algorithm as described in Sec. \ref{sec:fitting}.

\section{Variant model} \label{sec:variant_model}
To model the effect of new Omicron sub-variants, we used a simplified approach that can capture potential changes in intrinsic transmissibility and/or immune escape. This does not encompass the full dynamics of two or more variants spreading simultaneously with partial cross-immunity \cite{kucharski2016capturing}, but captured the key effects by changing relevant model parameters around a specified time point $t_{VOC}$ when the new variant (or mixutre of functionally equivalent variants) becomes dominant. For simplicity, we assume that all infections before $t_{VOC}$ are the resident variant and all infections after $t_{VOC}$ are the new variant (see Table \ref{tab:all_pars}). 

A variant that has different intrinsic transmissibility can be modelled by a change in the parameter $R_{EI}(t)$ at $t=t_{VOC}$. However, we assumed that the sub-variants considered in the simulation had the same intrinsic transmissibility as the first dominant variant BA.2, and derived their growth advantage solely from immune escape. 

A variant that evades vaccine-derived immunity can be modelled by reducing the initial antibody titre levels $e_{2d,0}$ and $e_{3d,0}$ for vaccinated but not previously infected states by an amount $\Delta n_{0,VOC}$ at $t=t_{VOC}$. This is equivalent to a reduction in vaccine effectiveness. 

Reducing the initial antibody titre for previously infected states ($k=11,\ldots,14$) would result in a permanent reduction in infected-induced immunity, including against future reinfection with the same variant. To avoid this, we instead modelled reduction in infection-derived antibody titre to the new variant by moving individuals in the previously infected states ($k=11$, $12$ or $13$) at $t=t_{VOC}$ to a lower immunity state ($k=12$, $13$ or $14$). This meant that a reduction in average titre was applied to people infected before $t=t_{VOC}$ (assumed to be infection with the resident variant), but people infected after $t=t_{VOC}$ (assumed to be infection with the new variant) started with the same initial antibody titre following recovery as before the new variant arrived. Thus, the model assumed an equally high level of homologous immunity against reinfection with the same variant, but a relatively lower level of cross-reactive immunity to reinfection with a new variant. 

We implemented the loss of immunity due to the new variant by applying a time-limited increase in the magnitude of the waning fluxes in Eq. (\ref{eq:waneTerm}) for the post-infection compartments:
\begin{equation} \label{eq:waneTermWithVOC}
W_{ik} = \left(r_w + r_{VOC} \phi\left(\frac{t-t_{VOC}}{\sigma_{VOC}} \right)\right) \left(\sum_{l=1}^{n_S} S_{il} Q^S_{lk}   + \hat{r} \sum_{l=1}^{n_S} R_{il} Q^R_{lk}\right), \qquad k=11,12,13,14
\end{equation}
where $\phi(.)$ is the standard normal probability density function. The magnitude of the drop in infection-derived immunity to the new variant was determined by a dimensionless parameter $r_{VOC}$ (see Table \ref{tab:all_pars}). Under this formulation, previously infected people moved to a lower post-infection immunity compartment in a short time window around $t=t_{VOC}$, of duration determined by the parameter $\sigma_{VOC}$. In the limit $\sigma_{VOC}\to0$, this movement would occur as an instantaneous pulse; larger values of $\sigma$ correspond to a more gradual change. We chose an arbitrary value of $\sigma_{VOC}=2$ days, representing a relatively rapid takeover of the new variant from the previously dominant variant; larger values of $\sigma_{VOC}$ would result in a more gradual change in the epidemic growth rate.

We assumed that the intrinsic disease severity was the same for all Omicron sub-variants (although changes in time-dependent realised severity could occur due the immune evasion process described above).

\section{Testing and clinical pathways}\label{sec:clinical_pathways}
The process of testing and progress to different clinical endpoints (hospital admission, hospital discharge, and death) can be modelled downstream of the transmission dynamics. We model the number of newly infectious people in each age group who will eventually become a confirmed case ($C$), be hospitalised ($H$), and die ($F$) via the differential equations.
\begin{eqnarray}
\frac{dC_{i1}}{dt} &=& 1/t_E \sum_{k=1}^{n_S}\left(p_\mathrm{test,clin,i}p_{\mathrm{clin},i}\frac{1-e_{S,k}}{1-e_{I,k}} + p_\mathrm{test,sub,i}\left(1-p_{\mathrm{clin},i}\frac{1-e_{S,k}}{1-e_{I,k}} \right) \right)E_{ik} \nonumber \\
&& - \alpha_1 C_{i1}  \\
\frac{dH_{i1}}{dt} &=& 1/t_E \mathrm{IHR}_i \sum_{k=1}^{n_S} \frac{1-e_{H,k}}{1-e_{I,k}} E_{ik} -  \alpha_1 H_{i1} \\ 
\frac{dF_{i1}}{dt} &=& 1/t_E \mathrm{IFR}_i \sum_{k=1}^{n_S} \frac{1-e_{F,k}}{1-e_{I,k}} E_{ik} -  \alpha_1 F_{i1}
\end{eqnarray}
where $\mathrm{IHR}_i$ and $\mathrm{IFR}_i$ are respectively the infection hospitalisation ratio and the infection fatality ratio for immune naive individuals in age group $i$ (see Table \ref{tab:age_dep_pars}).

The probability of testing positive for clinical and subclinical cases in age group $i$ is $p_\mathrm{test,clin,i}$ and $p_\mathrm{test,sub,i}$ respectively (see Table \ref{tab:all_pars}). We model a time- and age-dependent probability of testing, with different values of the testing probability $p_\mathrm{test,clin,i}$ in the three broad age bands: 0--30 years, 30--60 years and over 60 years. We assumed a constant value within each age band up to 30 April 2022, followed by a linear decline for the 0--30 years and 30--60 years age bands between the 1 May and 31 December 2022, followed by a constant value from 1 January 2023 onwards (Table \ref{tab:all_pars}). We assumed that the testing probability remains constant for over-60-year-olds. The values chosen for the testing probabilities and time window for their decline were manually tuned to reflect the observed age-structure of reported cases and the observed case hospitalisation ratio in each age group. We also apply a single fitted multiplier $\alpha_T \in U[0.8, 1.2]$ to all testing probabilities. The probability of testing positive for subclinical cases in age group $i$ is defined to be $p_\mathrm{test,sub,i} = 0.4p_\mathrm{test,clin,i}$.

\begin{figure}[p!]
    \centering
   \includegraphics[width=\linewidth]{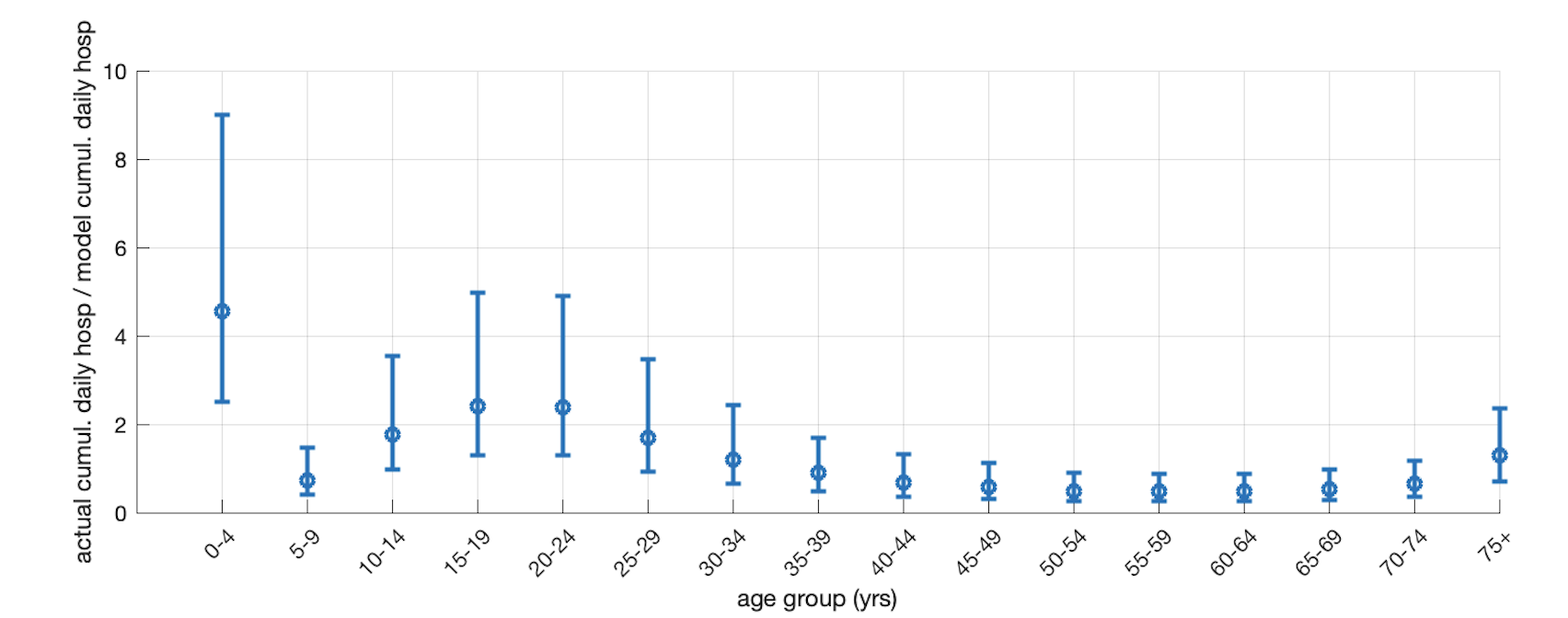}
    \caption{Ratio (median and 95\% CrI) of cumulative actual hospital admissions to cumulative modelled hospital admissions using a previous set of assumed values \cite{lustig2023modelling} for the age-dependent infection hospitalisation ratio. The model IHR was subsequently adjusted by multiplying by the median ratio in each age band resulting in the IHR values shown in Table \ref{tab:age_dep_pars} that were used in this study. Cumulative modelled hospital admissions were calculated for the period 25 January 2022 to 18 March 2023 from model simulations run on 22 March 2023, using parameters fitted to data up until 25 February 2023 and vaccination data until 13 February 2023.}
    \label{fig:IHRscalingfactors}
\end{figure}

The age distribution of hospitalisations in a previous version of this model had a significant discrepancy with the observed data \cite{lustig2023modelling}. This may have been because the age profile of immune naive IHR values assumed by the previous model was based on international data from pre-Omicron variants \cite{herrera2022age}.
To better model the observed age-distribution of hospitalisations in New Zealand during the Omicron waves, we calculated age-dependent adjustment factors to the previously assumed values of IHR. We did this by calculating the ratio of actual cumulative hospital admissions for the period 25 January 2022 to 18 March 2023 to modelled cumulative hospital admissions for the same period under the previous model \cite{lustig2023modelling} (see Figure \ref{fig:IHRscalingfactors}). These show that the previous model underestimated hospitalisations in the 0--5 years, 10--35 years, and over 75 years age groups, and overestimated hospitalisations in the 5--9 years and 40--75 years age groups. The IHR values shown in Table \ref{tab:age_dep_pars} are the adjusted immune naive IHR values used for the results shown in this study. Note that these IHR values are subsequently scaled by as single multiplier in each model run, which enables the overall IHR to be fitted to data using the ABC method (see Sec. \ref{sec:fitting}). We did not attempt to fit all 16 age-specific IHR values independently within the ABC routine as this would have resulted in too many free parameters.

The model also includes the effect of antiviral treatments on the infection-fatality rate (IFR) based on the proportion of cases prescribed antivirals in each age group. In New Zealand, eligiblity criteria for antiviral medications were widened in stages between July and September 2022. From September 2022 onwards, people aged over 65 years, M\=aori or Pacific people aged over 50 years, and other high-risk groups are eligible for free antivirals. This is one possible contributing factor to the observed reduction in age-specific case fatality ratios over time (after the modelled effect of vaccine and infection-derived immunity), consistent with an increase in antiviral eligibility and uptake over time. 

To account for this, we modelled the IFR for immune naive individuals in age group $i$ infected on day $t$ as a decreasing linear reduction of the proportion of cases prescribed antivirals:
\begin{equation} \label{eq:IFR_antivirals}
    \mathrm{IFR}_{\mathrm{post-antivirals}i}(t) = \mathrm{IFR}_i  \left(1-\frac{\alpha_A \hat{A}_i(t)}{\hat{C}_i(t)}\right),
\end{equation}
with $\hat{A}_i(t)$ being the number of cases in age group $i$ reported on day $t$ who received antivirals, $\hat{C}_i(t)$ being the number of reported cases in age group $i$ at time $t$, and $\alpha_A \sim U[0.4,0.6]$ being a fitted parameter quantifying the effect size (Table \ref{tab:all_pars}). This range of effect sizes is consistent with Ministry of Health analysis of the effect of Paxlovid and molnupiravir (the main two antiviral treatments used in New Zealand) on mortality risk \cite{MOH2023}. Note Eq. (\ref{eq:IFR_antivirals}) is the IFR for immune naive individuals. The IFR for individuals with vaccine-derived and/or infection-derived immunity is further reduced from Eq. (\ref{eq:IFR_antivirals}) according to the immunity model described in Sec. \ref{sec:immunity_model}. For scenarios (0a) and (1a) with no effect of antivirals, we set the parameter $\alpha_A$ to $0$.

\begin{figure}
      \centering
    \includegraphics[width=\textwidth, trim = 2cm 0 2cm 0, clip]{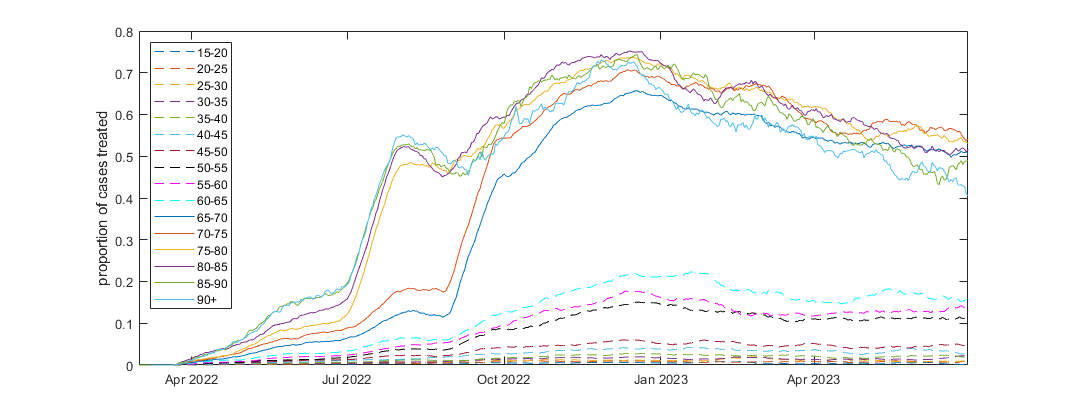}
    \caption{Proportion of reported cases that filled a prescription for either Paxlovid or molnupiravir within 7 days of report date or hospital admission date, calculated using a 28-day moving window.  }
    \label{fig:antivirals}  
\end{figure}

We used linked Ministry of Health prescribing data to calculate the number of cases  $\hat{A}_i(t)$ in age group $i$ reported on day $t$ who filled a prescription for either Paxlovid or molnupiravir within 7 days of report date or hospital admission date (see Figure \ref{fig:antivirals}). The most recent week of data were discarded to account for reporting lags. Data were then smoothed using a 8-week rolling average. The proportion of cases receiving antivirals $\hat{A}_i(t)/\hat{C}_i(t)$ was assumed to remain constant after the last date for which data was available. 

The effect of antiviral prescribing on the risk of hospitalisation is harder to estimate because in some cases antivirals are prescribed after onset of severe illness or hospital admission. We assumed there was no effect of antiviral treatments on the hospitalisation rate.

The time lag from onset of infectiousness to each endpoint was modelled via transition through a series of compartments:
\begin{equation} \label{eq:clinical_pathways}
\begin{array}{lll}
\frac{dC_{i,2}}{dt} = \alpha_{1} C_{i1} - \alpha_{2} C_{i2}, \quad & 
\frac{dH_{i,2}}{dt} = \alpha_{1} H_{i1} - \alpha_{2} H_{i2}, \quad &
\frac{dF_{i,2}}{dt} = \alpha_{1} F_{i1} - \alpha_{2} F_{i2}, \\
\frac{dC_{i,3}}{dt} = \alpha_{2} C_{i2}, \quad & 
\frac{dH_{i,3}}{dt} = \alpha_{2} H_{i2} - \alpha_{3} H_{i3}, \quad &
\frac{dF_{i,3}}{dt} = \alpha_{2} F_{i2} - \alpha_{3} F_{i3}, \\
&
\frac{dH_{i,4}}{dt} = \alpha_{3} H_{i3} - \alpha_{4,i} H_{i4}, \quad &
\frac{dF_{i,4}}{dt} = \alpha_{3} F_{i3} - \alpha_{4}' F_{i4}, \\
&
\frac{dH_{i,5}}{dt} = \alpha_{4,i} H_{i4}, \quad &
\frac{dF_{i,5}}{dt} = \alpha_{4}' F_{i4} - \alpha_{5} F_{i5}, \\
&
&
\frac{dF_{i,6}}{dt} = \alpha_{5} F_{i5}.
\end{array}
\end{equation}
where $\alpha_k$ are a set of rate constants determining the time lags. We set $\alpha_1=\alpha_2=2/t_T$ where $t_T$ is the mean time from onset of infectiousness to return of a positive test result. The mean time from positive test result to hospital admission is $t_H=\alpha_3^{-1}$, and the mean length of hospital stay for non-fatal cases in age group $i$ is $t_{LOS,i}=\alpha_{4,i}^{-1}$. We set $\alpha_4'=\alpha_5=2/t_F$ where $t_F$ is the mean time from hospital admission to death. 

The compartment $C_{i3}$ represents the observed cumulative number of cases, $H_{i4}$ the number of cases currently in hospital, $H_{i5}$ the cumulative number of hospital discharges and $F_{i6}$ the cumulative number of fatalities in age group $i$ at time $t$. The other $C$, $H$ and $F$ variables above represent latent (unobservable) states.

\section{Fitting to data} \label{sec:fitting}

The variables in Eqs. (\ref{eq:clinical_pathways}) were used to define a number of key model outputs for model fitting and/or comparison with data:
\begin{enumerate}
\item Total new cases per day: $\alpha_2\sum_i C_{i2}(t)$
\item Proportion of new cases in age group $i$: $C_{i2}(t)/\sum_j C_{j2}(t)$
\item Total new admissions per day: $\alpha_3 \sum_i H_{i3}(t)$
\item Proportion of new admission in age group $i$: $H_{i3}(t)/\sum_j H_{j3}(t)$
\item New deaths per day: $\alpha_5 \sum_i F_{i5}(t)$
\item New infections per capita per day: $1/t_E \sum_{i,k} E_{ik}(t)/\sum_i N_i(t)$
\end{enumerate}

Outputs (1)--(2) were fitted to data from the Ministry of Health on total and age-stratified new daily Covid-19 cases smoothed using a 7-day rolling average. The start date of 1 March 2022 was chosen to avoid using data from a period at the start of the first Omicron wave when case ascertainment was likely significantly lower due to a lack of testing availability. Ten year age bands were used for age stratification.

Output (3)--(4) were fitted to data on total and age-stratified new daily Covid-19 hospital admissions from 1 February 2022, smoothed using a 7-day rolling average. Only hospital admissions categorised by the Ministry of Health as a ``Covid-related hospitalisation'' were included.  Ten year age bands were used for age stratification.

Output (5) was fitted to daily Covid-19 deaths 1 February 2022, smoothed using a 7-day rolling average. Deaths were defined to be cases that were recording as having died and where the cause-of-death summary was ``COVID as underlying'', ``COVID as contributory'', or ``Not available''; deaths where the cause-of-death summary was ``Not COVID'' were excluded. 

Output (6) was fitted to data on the weekly incidence of new cases per 1000 people in a routinely tested cohort of approximately 20,000 border workers from 13 February to 3 July 2022. This may not be a representative sample of the population but we included it because, unlike outputs (1--5), it provides longitudinal surveillance data that is less sensitive to either case ascertainment levels or disease severity.

For each fitted time series (1), (3), (5) and  (6), we defined the error function as 
\begin{equation}\label{eq:errorFunc}
d(x,y) = \frac{1}{n} \sum_{t=1}^n  \left(\ln (x_t+\epsilon) - \ln(y_t+\epsilon) \right)^2,
\end{equation}
where $x_t$ and $y_t$ are the model output and data respectively for day $t$, and $\epsilon$ is a fixed value that is small relative to typical values of the variable being fitted. The error function for outputs (2) and (4) was defined similarly, but with the summation being over age bands $i$ as well as time $t$. We set $\epsilon=1$ for outputs (1), (3) and (5), $\epsilon=0.01$ for outputs (2) and (4), and $\epsilon=10^{-4}$ for output (6). 

The total model error was defined as the sum of the error for outputs (1)--(6). To implement ABC rejection, we solved the model for $N=15,000$ parameter combinations drawn randomly from the prior and retained the 1\% of simulations with the smallest error. We report 95\% credible intervals (CrI) for each model output across the retained simulations. For time series graphs, we display the 95\% curvewise credible interval (CrI), i.e. the envelope that contains the curves corresponding to the best-fitting 95\% of accepted simulations across the simulated time period. This is a better indicator of model fit and model uncertainty than using fixed-time statistics, such as the interval containing 95\% of accepted simulations at each time $t$ \cite{juul2021fixed}.

\section{Mortality and hospitalisation rates by ethnicity}
\label{sec:ethnicity_rates}

In order to compare Covid-19 mortality and hospitalisation rates across ethnic groups, we extracted the number of deaths where the cause of death was classified as ``COVID underlying'' or ``COVID contributory'' and the number of hospital admissions classified as Covid-19-related in the following ethnicities: M\=aori, Pacific Peoples, Asian, European/other. Prioritised ethnicity data was available. 

M\=aori and Pacific populations have a substantially younger age structure compared to the European population, so we calculated age-standardised rates. For population denominators, we used the 2022 HSU population \cite{statsnz_hsu}. Population size data was available in the same ethnic groups as above and in five-year age bands up to 90 years and over.  
Age-standardised Covid-19 mortality and hospitalisation rates for the period from 1 January 2022 and 30 June 2023 are shown in Table \ref{tab:ethnicity_rates}.

\afterpage{

\thispagestyle{empty}

\vspace{-5cm}

\begin{table}[!ht]
\addcontentsline{toc}{section}{Supplementary Tables and Figures}
\fontsize{8}{9}\selectfont
\hspace{-0.6cm}
\begin{tabular}{p{9.7cm}lp{2.1cm}}
\hline
\textbf{Parameter} & \textbf{Value} & \textbf{Source}\\
\hline
\textit{Epidemiological parameters} & \\
\hline
Latent period         &  $t_E=1$ day & \cite{wu2022incubation}  \\
Infectious period         &  $t_I=2.3$ days & \cite{abbott2022estimation}  \\
Mean time from onset of infectiousness to positive test result & $t_T=4$ days & \cite{vattiato2022assessment} \\
Mean time from test result to hospital admission & $t_H=1$ days & Assumed \\
Mean time from admission to death  & $t_F=14$ days & Assumed \\
Relative infectiousness of subclinical individuals & $\tau=0.5$ & \cite{davies2020age} \\
\hline
\textit{Date-specific parameters} & \\
\hline
Date of seeding with infectious cases & 19 Jan 2022 $ +U[-3,3]$ & Fitted \\
Number of seed cases in age group $i$ & $0.0001N_i$ & Assumed \\
$R_{EI}(t)$ in period 1 & $R_{EI,1}\sim U[1.9,2.5]$ & Fitted \\
$R_{EI}(t)$ in period 2 & $R_{EI,2}\sim U[2.9,4.3]$ & Fitted \\
$R_{EI}(t)$ in period 3 & $R_{EI,2}* U[1.1, 1.3]$ & Fitted \\
End of period 1 & 10 Mar 2022 $+U[-5,5]$ & Fitted \\
End of period 2 & 15 Sep 2022 $+U[-5,5]$ & Fitted \\
Period 1 -- period 2 ramp window & $U[35,75]$ days & Fitted \\
Period 2 -- period 3 ramp window & $U[1,19]$ days & Fitted \\
Relaxation of contact matrix & $\alpha_M\sim U[0,0.8] $ & Fitted \\
Contact matrix ramp window &  $U[50,90]$ days & Fitted \\
Testing prob. before 1 May 2022 (clinical, 0-30 yrs) &  $p_\mathrm{test1,clin,0-30}=0.5$ & Manually tuned \\
Testing prob. after 1 Jan 2023 (clinical, 0-30 yrs) &  $p_\mathrm{test2,clin,0-30}=0.25$ & Manually tuned \\
Testing prob. before 1 May 2022 (clinical, 30-60 yrs) &  $p_\mathrm{test1,clin,30-60}=0.6$ & Manually tuned \\
Testing prob. after 1 Jan 2023 (clinical, 30-60 yrs) &  $p_\mathrm{test2,clin,30-60}=0.4$ & Manually tuned \\
Testing prob. (clinical, 60+ yrs) &  $p_\mathrm{test,clin,60+}=0.75$ & Manually tuned \\
Testing prob. (subclinical, age group $i$) & $p_\mathrm{test,sub,i}=0.4p_\mathrm{test,clin,i}$ & Assumed \\
Testing prob. global multiplier & $\alpha_T \sim U[0.8, 1.2]$ & Fitted \\
\hline
\textit{Variant model} & \\
\hline
BA.5 escape from infection-derived immunity & $r_{VOC1}\sim U[0.1, 0.7]$ & Fitted \\
BA.5 change in vaccine-derived log antibody titre relative to BA.2 & $\Delta n_{0,VOC1}= -0.92$ & \cite{khan2022omicron} \cite{hachmann2022neutralization} \\
BA.5 dominance date & $t_{VOC1}=$ 20 Jun 2022 & Manually tuned \\
CH.1.1/BQ.1.1 escape from infection-derived immunity & $r_{VOC2}=0.25$ & Manually tuned \\
CH.1.1/BQ.1.1 change in vaccine-derived log antibody titre relative to BA.5 & $\Delta n_{0,VOC2}= 0$ & Assumed \\
CH.1.1/BQ.1.1 dominance date & $t_{VOC2}=$ 15 Nov 2022 & Manually tuned \\
Variant transition window & $\sigma_{VOC}=2$ days & Assumed \\
\hline
\textit{Immunity model} & \\
\hline
Initial log antibody titre &&  \\
    - 2 doses  &  $n_{2d,0}=-1.61$ & \cite{golding2022analyses}  \\
    - 3 doses  &  $n_{3d,0}=-0.92$ & \cite{golding2022analyses} \\
    - prior infection with 0/1 doses  & $n_{p,0}=1.39$ & Manually tuned \\
    - prior infection with 2 doses    & $n_{p2d,0}=2.71$ & Manually tuned  \\
    - prior infection with 3 doses    & $n_{p3d,0}=3.56$ & Manually tuned \\
Log antibody titre providing 50\% immunity: &&    \\
 - against infection & $n_{\mathrm{inf},50}=-1.61$ & \cite{khoury2021neutralizing}  \\
 - against hospitalisation & $n_{\mathrm{hosp},50}=-3.51$ & \cite{khoury2021neutralizing} \\
 - against death & $n_{\mathrm{death},50}=-3.51$ &\cite{khoury2021neutralizing} \\
Waning rate & $r_w\sim U[0.0027,0.0063]$ day$^{-1}$ & Fitted \\
Relative rate of moving from $R$ to $S$ & $\hat{r}=1.85$ & Assumed \\
Drop in log titre in subsequent compartment & $n_\mathrm{drop}=2.30$ & Assumed \\
Slope of logistic function  & $\kappa=1.28$ & \cite{khoury2021neutralizing} \\
Minimum long-term immunity to hospitalisation and death & $e_\mathrm{sev,min}=0.5$ & Assumed\\
Global IHR multiplier & $\alpha_\mathrm{IHR}\sim U[0.5, 1.5]$ & Fitted\\
Global IFR multiplier & $\alpha_\mathrm{IFR}\sim U[0.5, 1.5]$ & Fitted\\
Antiviral effect on IFR & $\alpha_A \sim U[0.4, 0.6]$ & Fitted\\
\hline
\end{tabular}
\caption{Model parameter values and prior distributions.}
\label{tab:all_pars}
\end{table}
\clearpage
}

\newpage 

\begin{table}[p]
\centering
    \small
\begin{tabular}{p{1.5cm}p{1.5cm}p{1cm}p{1cm}p{1.5cm}p{1.5cm}p{1cm}p{2cm}}
\hline
Age (yrs) & Popn $N_i(0)$ & $u_i$ & $p_{\mathrm{clin},i}$ & $\mathrm{IHR}_i$ per 1000 & $\mathrm{IFR}_i$ per 1000 & $t_{LOS,i}$ (days) & $\mu_i$ (per 1000 per yr) \\
\hline
0-5  & 310660   & 0.46 & 54\% & 7.12  & 0.0034 & 2.0 & 1.07 \\
 5-10 & 332944   & 0.46 & 55\% & 1.15  & 0.0034 & 2.0 & 0.08 \\
 10-15 & 345479 & 0.45 & 58\% & 1.19  & 0.0034 & 2.0 & 0.17 \\
 15-20 & 318994 & 0.56 & 60\% & 2.42  & 0.0062 & 2.0 & 0.41\\
 20-25 & 334793 & 0.79 & 62\% & 3.48  & 0.012 & 2.0 & 0.60 \\
 25-30 & 382666 & 0.93 & 64\% & 3.54  & 0.024 & 2.0 & 0.56 \\
 30-35 & 404515 & 0.97 & 66\% & 3.69  & 0.048 & 2.7 & 0.73\\
 35-40 & 358897 & 0.98 & 68\% & 3.99  & 0.091  & 3.3 & 0.83 \\
 40-45 & 324200 & 0.94 & 70\% & 4.36  & 0.180  & 4.0 & 1.21\\
 45-50 & 325381 & 0.93 & 71\% & 5.49  & 0.360  & 4.7 & 1.95\\
 50-55 & 336644 & 0.94 & 73\% & 6.68  & 0.697  & 5.4 & 3.07\\
 55-60 & 324299 & 0.97 & 74\% & 9.18  & 1.35  & 6.0 & 4.45\\
 60-65 & 303456 & 1.00 & 76\% & 13.48  & 2.65  & 6.7 & 6.49\\
 65-70 & 258073 & 0.98 & 77\% & 21.44  & 5.08  & 7.4 & 10.27\\
 70-75 & 220811 & 0.90 & 78\% & 36.31  & 9.74   & 8.0 & 16.69\\ 
 75+ & 351834   & 0.86 & 80\% & 130.03  & 54.7   & 8.7 & 136.0\\ 
\hline
\end{tabular}
    \caption{Age-dependent model parameters. `Popn' is the initial population size in each age group, as of October 2022; $u_i$ is the susceptibility of age group $i$ relative to the 60-65 year age group; $p_{clin,i}$, $\mathrm{IHR}_i$ and $\mathrm{IFR}_i$ are respectively the proportion of infections causing clinical disease, hospitalisation and death respectively for individuals with no immunity (i.e. unvaccinated and no prior infection); $t_{LOS,i}$ is the average length of hospital stay estimated from MOH data on duration of patients receiving hospital treatment for Covid-19; $\mu_i$ is the all-cause death rate per 1000 people per year. Values of $p_{clin,i}$ are from \cite{hinch2021openabm}. The age-dependence in $\mathrm{IFR}_i$ is based on the results of \cite{herrera2022age} but values are scaled down for consistency with observed death rates, reflecting reduced virulence of Omicron relative to earlier variants. Values of $\mathrm{IHR}_i$ were set as described in section \ref{sec:clinical_pathways}. The values of $\mathrm{IHR}_i$ and $\mathrm{IFR}_i$ in the Table are the mean of the prior distribution for these parameters. In each model simulation, the vectors of $\mathrm{IHR}_i$ values and $\mathrm{IFR}_i$ values are multiplied by  fitted global adjustment factors $\alpha_\mathrm{IHR}$ and $\alpha_\mathrm{IFR}\sim U[0.5,1.5]$ (Table \ref{tab:all_pars}). Total birth rate assumed to be $b=59637$ yr$^{-1}$.}
\label{tab:age_dep_pars}
\end{table}

\newpage

\begin{table}
\centering
\begin{tabular}{p{8.1cm}p{7.8cm}}
\hline
Scenario type & Definition \\
\hline
A. Vaccination rates reduced by factor of $p_v$ & $v_{id}(t)=p_v \hat{v}_{id}(t) $  \\
B. No vaccination in age groups $1$ to $k$ & $v_{id}(t)=0$ for $i=1,\ldots,k$ and $v_{id}(t)=\hat{v}_{id}(t)$ for $i=k+1,\ldots,16$ \\
C. Adult vaccination rates as for age group $k$ & $v_{id}(t)=\hat{v}_{kd}(t) N_i/N_k$ for $i=4,\ldots,12$ and $v_{id}(t)=\hat{v}_{id}(t)$ for $i=1,\ldots,3$   \\
D. Vaccination rates as for ethnic group $G$ & $v_{id}(t)=\hat{v}^{(G)}_{id}(t) N_i/N^{(G)}_i $ \\
\hline
\end{tabular}
\caption{Definition of the counterfactual scenarios considered. Each scenario is defined by the number of $d^\mathrm{th}$ does given in age group $i$ at time $t$, denoted $v_{id}(t)$, which is set relative to the actual number of doses, denoted $\hat{v}_{id}(t)$. Four types of method (A--D) are used for setting $v_{id}(t)$. Scenarios (0), (1) and (2) in Table 1 are of Type A with $p_v=1$, $p_v=0$ and $p_v=0.9$ respectively. Scenario (3) is of Type B with $k=12$ (representing the 55-60-year-old group). Scenario (4) is of Type C and sets $v_{id}(t)$ so that the number of vaccine doses per capita in age group $i$, $v_{id}(t)/N_i$, is equal to the actual number of doses per capita in age group $k$, $\hat{v}_{kd}(t)/N_k$, with $k=5$ (representing 20-25-year-olds). Scenarios (5) and (6) are of Type D and set $v_{id}(t)$ so that the number of vaccine doses per capita in age group $i$, $v_{id}(t)/N_i$ is equal to the actual number of doses per capita in age group $i$ for ethnic group $G$. Here $N_i$ and $N^{(G)}_i$ denote the number of people in age group $i$ at time $t=0$ in the whole population and in ethnic group $G$ respectively. }
\label{tab:counterfactuals}
\end{table}

\begin{table}
\centering
\begin{tabular}{lll}
\hline
& Mortality rate & Hospitalisation rate \\
\hline
M\=aori &                         $100$  & $850$ \\
Pacific Peoples &                 $109$  & $1190$ \\
Asian &                           $38$  & $460$ \\
European/other &               $58$  & $490$  \\
\hline
Overall &                         $61$ & $550$ \\
\hline
\end{tabular}
\caption{Age-standardised Covid-19 mortality and hospitalisation rates per 100,000 people by prioritised ethnicity for the period 1 January 2022 to 30 June 2023. Covid-19 deaths were defined to be those where the cause of death summary was either `COVID as underlying'' or ``COVID as contributory''; Covid-19 hospitalisations were defined to be those  Rates are standardised to the total New Zealand population according to the HSU dataset in 2022.  }
\label{tab:ethnicity_rates}
\end{table}

\newpage

\begin{landscape}
\begin{table}
\small
    \centering
    \begin{tabular}{p{3.9cm}>{\raggedright\arraybackslash}p{3.0cm}>{\raggedright\arraybackslash}p{2.7cm}>{\raggedright\arraybackslash}p{3.1cm}>{\raggedright\arraybackslash}p{2.8cm}}
    \hline

     Scenario    & $\Delta$ infections (millions) & $\Delta$ admissions  (thousands)  & $\Delta$ deaths  & $\Delta$ YLL  (thousands)  \\
\hline 
(0) Baseline & - & - & - & -   \\ 
(1) No vaccine & 1.45 [1.33, 1.55] & 45.1 [34.4, 55.6] & 6650 [4424, 10180] & 74.5 [51.0, 115.4]   \\ 
(2) 10\% drop in rates & 0.13 [0.12, 0.14] & 4.2 [3.2, 5.2] & 603 [395, 924] & 6.8 [4.6, 10.7]   \\ 
(3) No vaccine in U60s & 1.19 [1.11, 1.25] & 14.5 [10.9, 17.1] & 773 [526, 1144] & 20.2 [13.5, 30.4]   \\ 
(4) 20-25-year-old rates & 0.20 [0.15, 0.24] & 4.2 [3.1, 5.4] & 733 [491, 1098] & 7.2 [4.8, 10.7]   \\ 
(5) Euro/other rates & -0.03 [-0.04, -0.03] & -0.5 [-0.6, -0.4] & -64 [-99, -42] & -0.6 [-1.0, -0.4]   \\ 
\hline 
(0a) No AVs & - & - & - & -   \\ 
(1a) No vaccine or AVs & 1.45 [1.33, 1.55] & 45.1 [34.4, 55.6] & 7604 [5080, 11942] & 82.4 [56.1, 129.3]   \\ 
\hline 
(5) Euro/other rates & - & - & - & -   \\ 
(6) M\=aori rates & 0.26 [0.23, 0.28] & 3.6 [2.7, 4.3] & 419 [284, 639] & 4.6 [3.1, 7.0]   \\ 

    \end{tabular}
    \caption{\small Model results (median and 95\% CrI) in each scenario for the difference ($\Delta$) in the total number of infections, hospital admissions, deaths, and years of life lost (YLL), between 1 January 2022 and 30 June 2023 relative to a comparison scenario. Scenarios (1)--(5) are compared against scenario (0) which is the baseline (factual) scenario. Scenario (1a) with no vaccination and no antivirals is compared against scenario (0a) with no antivirals (and actual vaccination rates). Scenario (6) with M\=aori vaccination rates is compared against scenario (5) with European/other vaccination rates. Note: differences ($\Delta$) are always calculated between two scenarios run with the same set of fitted parameters $\{\theta_i\}$;   results in the Table show the median and 95\% CrI of $\Delta$ across the 150 accepted parameter combinations.   }
    \label{tab:deltas}
\end{table}

\begin{table}
\small
    \centering
    \begin{tabular}{p{4.1cm}p{2.8cm}p{2.9cm}p{3.2cm}p{3cm}p{3cm}}
    \hline
     Scenario    & Infections \hspace{5mm} (millions) & Admissions \hspace{5mm} (thousands)  & Deaths  & YLL \hspace{12mm} (thousands) & Peak occupancy \\
     \hline
{\em Actual}  & -  & $28.8$ & $3196$   & $33.9^*$ & $1016^{**}$   \\
     \hline
{\em Model scenarios} \\
(0) Baseline & 5.67 [4.96, 6.59] & 28.8 [22.1, 36.3] & 3217 [2221, 4466] & 39.7 [27.2, 54.4] & 859 [573, 1299]  \\ 
(0a) No AVs & 5.67 [4.96, 6.59] & 28.8 [22.1, 36.3] & 3901 [2617, 5298] & 45.4 [31.1, 62.0] & 859 [573, 1299]  \\ 
(1) No vaccine & 7.11 [6.28, 8.12] & 74.2 [56.1, 91.4] & 9657 [6637, 13396] & 112.0 [77.6, 154.8] & 5583 [4041, 7319]  \\ 
(1a) No vaccine or AVs & 7.11 [6.28, 8.12] & 74.2 [56.1, 91.4] & 11146 [7639, 15813] & 125.7 [85.5, 175.8] & 5583 [4041, 7319]  \\ 
(2) 10\% drop in rates & 5.81 [5.08, 6.73] & 33.0 [25.1, 41.4] & 3799 [2635, 5282] & 46.3 [31.7, 63.3] & 1051 [717, 1428]  \\ 
(3) No vaccine in U60s & 6.86 [6.06, 7.82] & 43.1 [32.9, 53.2] & 3988 [2748, 5488] & 58.8 [40.8, 80.6] & 2877 [2096, 3704]  \\ 
(4) 20-25-year-old rates & 5.87 [5.09, 6.81] & 31.9 [24.3, 40.3] & 3684 [2551, 5136] & 44.6 [30.5, 61.2] & 879 [600, 1305]  \\ 
(5) Euro/other rates & 5.59 [4.88, 6.50] & 27.5 [21.1, 34.6] & 3082 [2126, 4270] & 38.0 [26.1, 52.1] & 789 [507, 1258]  \\ 
(6) M\=aori rates & 6.04 [5.29, 6.97] & 37.0 [28.1, 46.1] & 4301 [2991, 5981] & 51.1 [35.0, 69.8] & 1368 [973, 1841]  \\ 
    \hline
    \end{tabular}
    \caption{\small Model results (median and 95\% CrI) in each scenario for the total number of infections, hospital admissions, deaths, and years of life lost (YLL), and the peak hospital occupancy, between 1 January 2022 and 30 June 2023, in the sensitivity analysis using StatsNZ population projections instead of HSU population data (see Section \ref{sec:pop_data}). Scenarios are: (0) baseline (actual vaccination rates); (0a) no antivirals (actual vaccination rates); (1) no vaccination; (1a) no vaccination or antivirals; (2) vaccination rates set to a proportion $p_v=0.9$ of actual vaccination rates at all ages; (3) no vaccination of under-60-year-olds; (4) vaccination rates at all ages set to actual vaccination rates in the 20-25-year-old group; (5-6) vaccination rates set to actual vaccination rates for European/other and M\=aori ethnicities respectively at all ages. $^*$Estimated using cohort life tables via the same method as for model YLL calculations. $^{**}$Includes some incidental hospitalisations (i.e. patients who were positive for Covid-19 but not receiving treatment for Covid-19).  }
    \label{tab:sensitivity_results}
\end{table}
\end{landscape}

\newpage

\begin{figure}[p!]
    \centering
    \includegraphics[width=1.0\textwidth, trim = 2cm 0 2cm 0, clip]{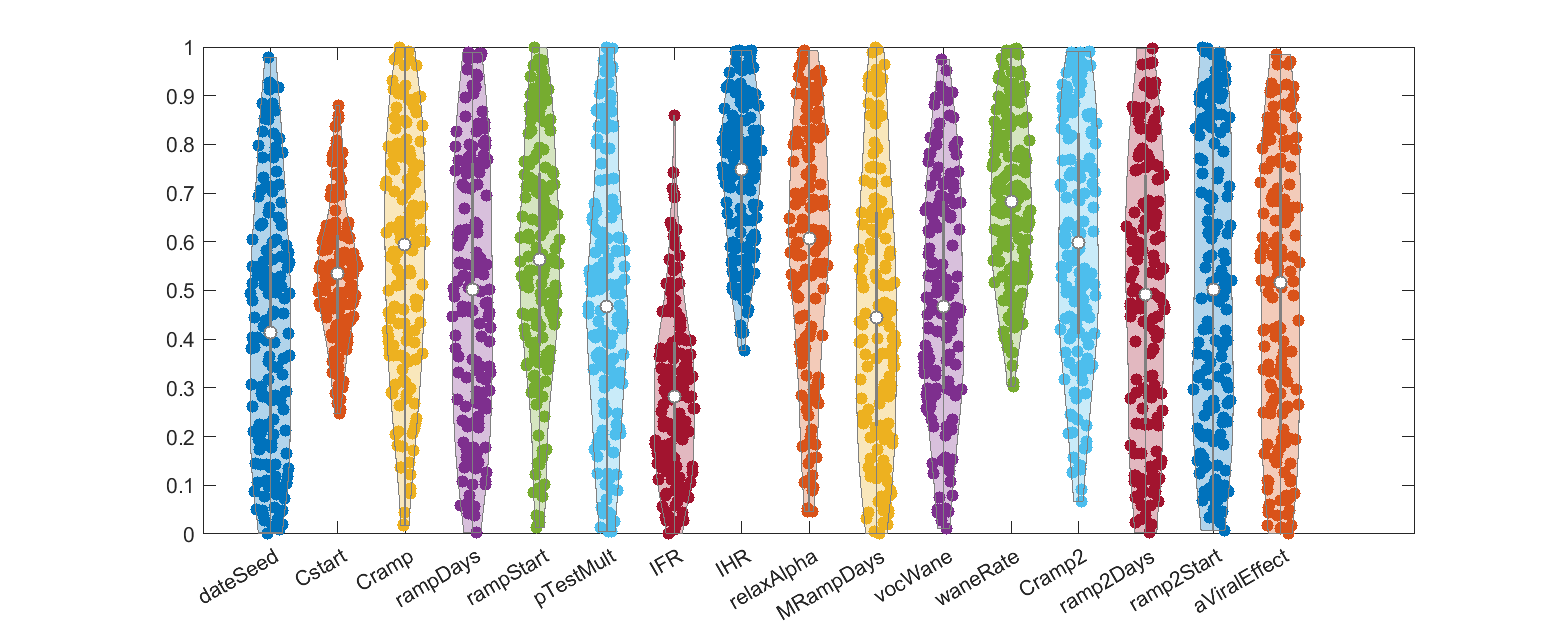}
    \caption[Posterior distribution of fitted parameters]{Violin plots showing the approximate marginal posterior distributions of each fitted parameter across the 150 accepted realisations of the model with the best fit to the data out of 15,000 random draws from the prior. Each parameter $\theta_i$ has a uniform prior $\theta_i\sim U[a_i,b_i]$ (see Table \ref{tab:all_pars}) and for the purposes of plotting, each parameter is transformed to the $[0,1]$ scale via $z_i=(\theta_i-a_i)/(b_i-a_i)$. }
    \label{fig:violin}
\end{figure}

\begin{figure}[p!]
    \centering
\includegraphics[width=\textwidth]{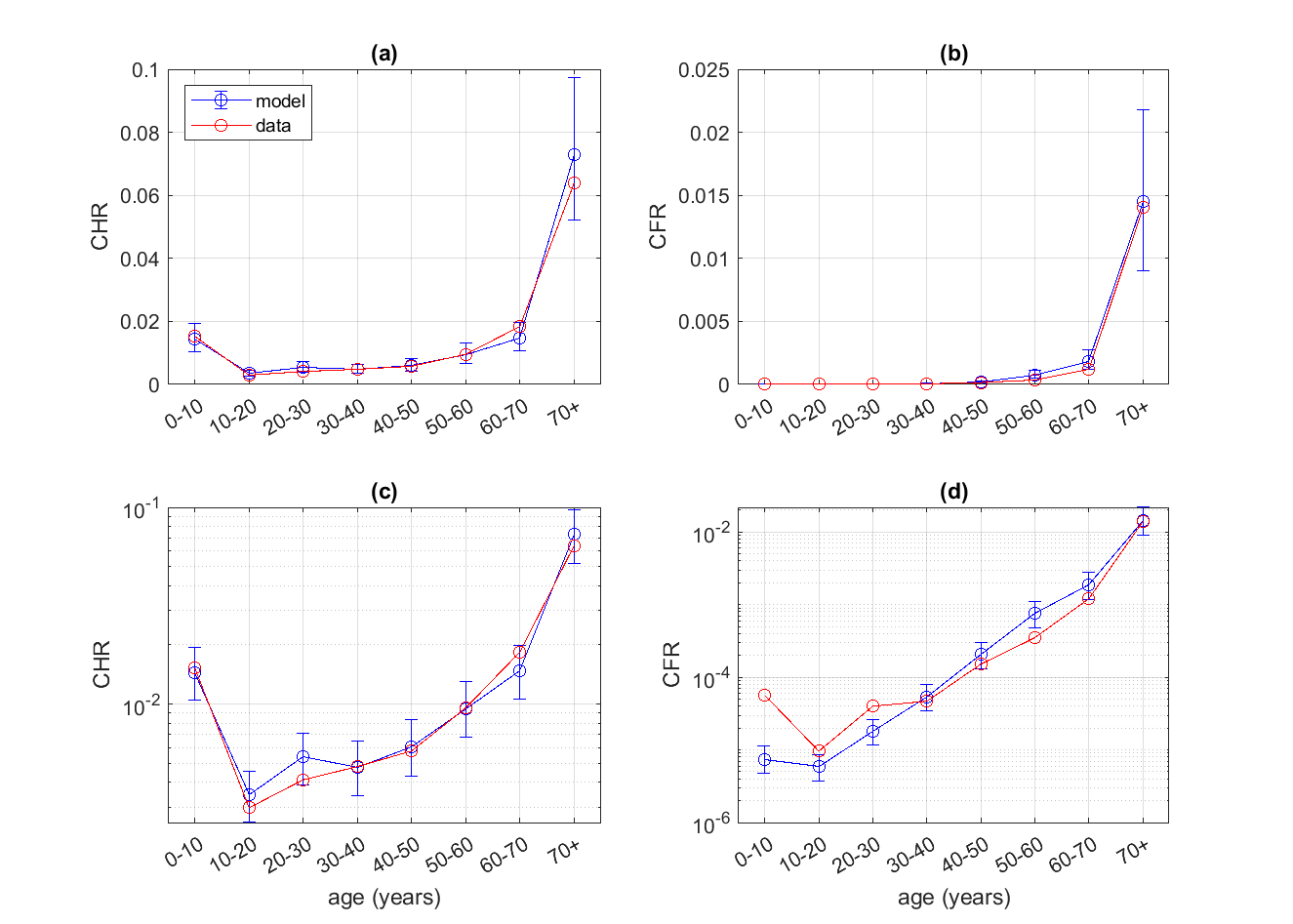}
    \caption[Age-specific case hospitalisation and case fatality ratios]{Age-specific case hospitalisation ratio (CHR) and case fatality ratio (CFR) between for model output in the baseline scenario (blue circles show median and error bars show 95\% CrI for the accepted model realisations) and data (red), for the period 25 January 2022 to 30 June 2023. Upper plots show results on a linear scale; lower plots show results on a log scale. }
    \label{fig:CHR_CFR}
\end{figure}

\begin{landscape}
\begin{figure}
    \centering
    \includegraphics[width=\linewidth,trim={2cm 0cm 3.5cm 0cm},clip]{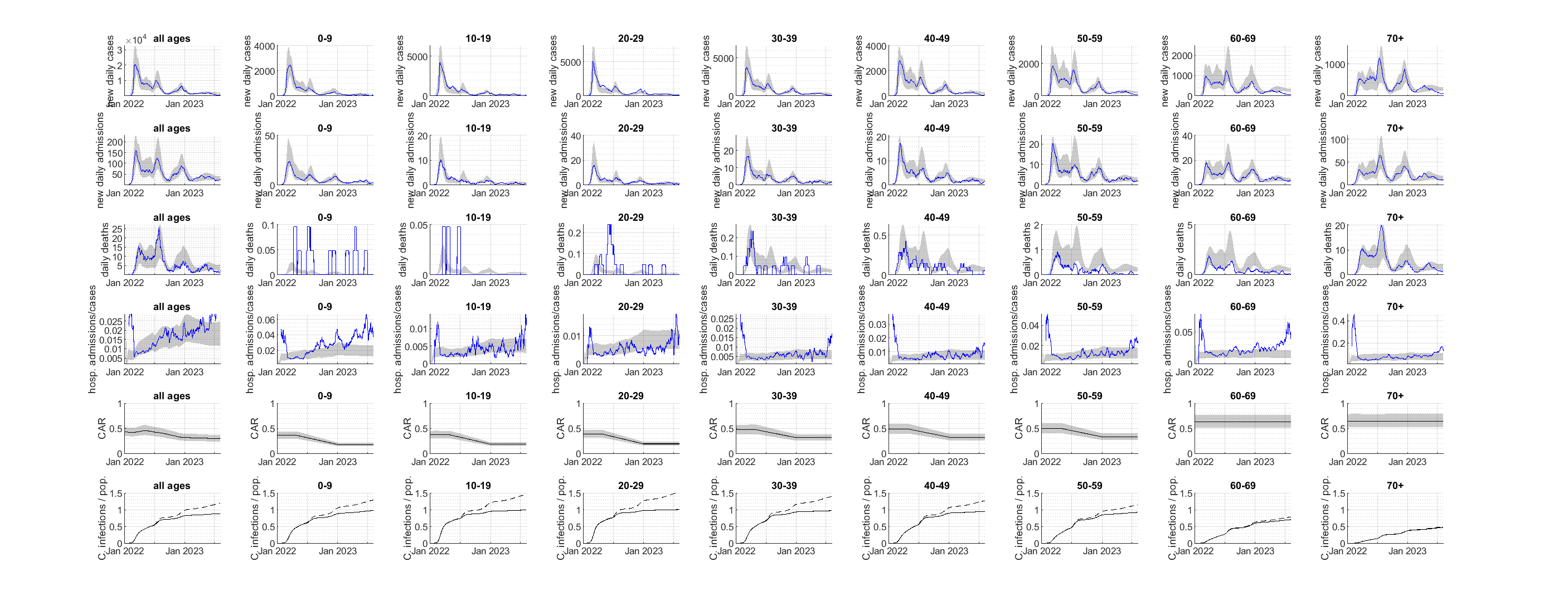}
     \caption{Age-stratified results for the baseline (factual) scenario showing: new daily cases; new daily hospital admissions; daily deaths; ratio of admissions to cases; ratio of cases to infections (case ascertainment rate, CAR); and cumulative infections relative to population size (solid line is first infections only; dashed line is all infections). Graphs show the curvewise 95\% credible interval (grey shaded area). Model was fitted to data (blue) up to 13 August 2023.  Note: model results were calculated in five-year age bands but plotted in ten-year age bands for ease of display. 
}
    \label{fig:modelresults_age}
\end{figure}
\end{landscape}

\begin{figure}
    \centering
    \includegraphics[trim={0 0 0 1.05cm},clip,width=\linewidth]{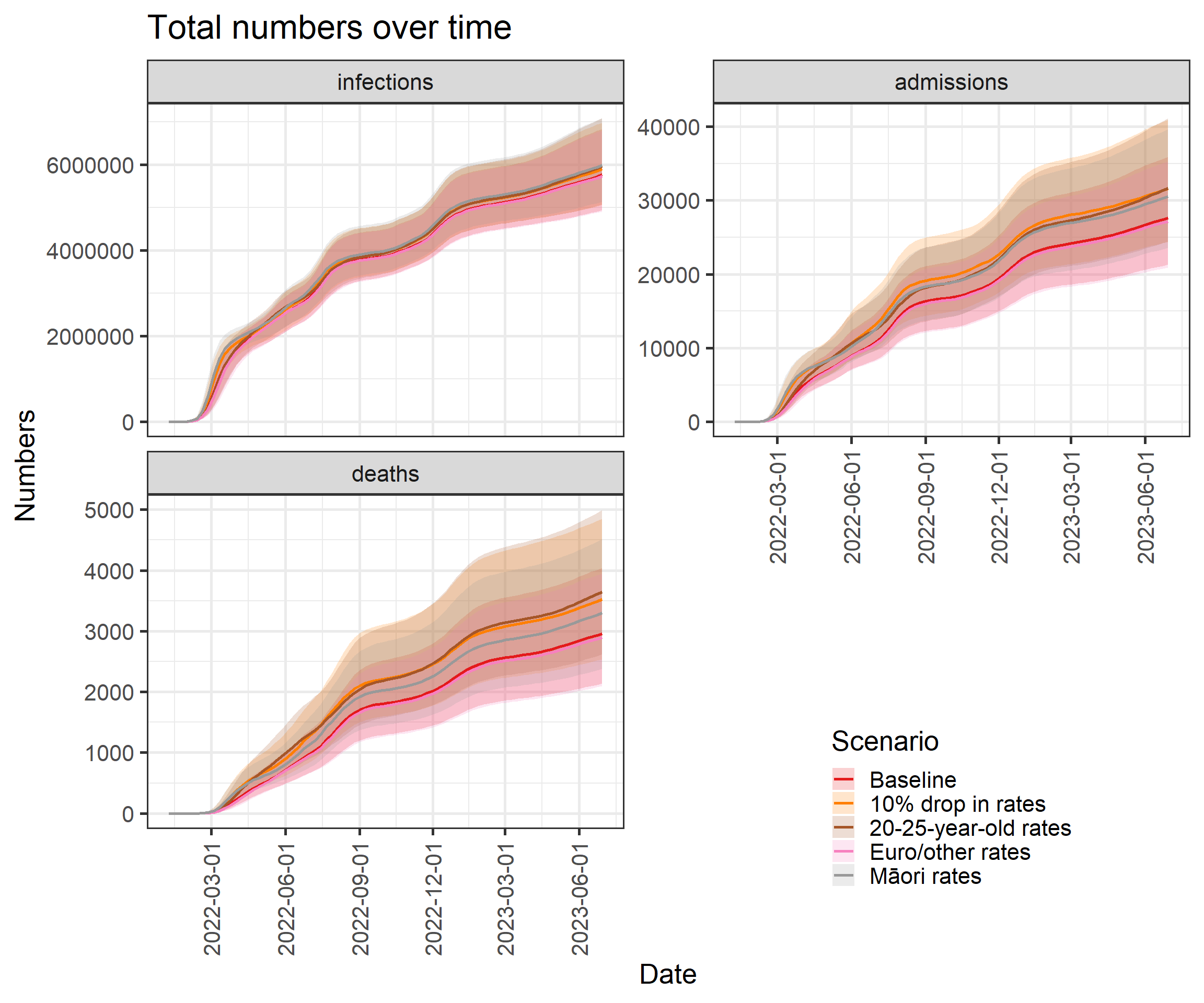}
    \caption{Cumulative number of infections, hospital admissions and deaths over time in the baseline scenario (red) and the scenarios with a 10\% reduction in vaccination rates (orange), vaccination rates for all adults set to actual vaccination rates in the 20-25-year-old group (brown), vaccination rates set to actual vaccination rates for European/other (pink) and M\=aori (grey). Graphs show the median (solid curves) and 95\% CrI  (shaded areas) for each scenario. See Figure 4 of the main article for corresponding results for the other scenarios considered. }
        \label{fig:cumulative}
\end{figure}

\clearpage

\addcontentsline{toc}{section}{Supplementary References}

\printbibliography